\documentclass[a4paper,UKenglish]{mylipics-v2018}

\usepackage{amsmath,amssymb,tabularx,hyperref,lineno}

\usepackage[end]{algpseudocode}
\usepackage{xspace,todonotes}

\long\def\ignore#1{\vskip 0pt}

\ignore{
\theoremstyle{plain}
\newtheorem{theorem}{Theorem}
\newtheorem{lemma}[theorem]{Lemma}
\newtheorem{corollary}[theorem]{Corollary}
\theoremstyle{definition}
\newtheorem{definition}[theorem]{Definition}

\newtheorem{property}[theorem]{Property}
\theoremstyle{remark}

}

\def\Oh{\mathcal{O}}

\newcommand{\xx}{\$}
\newcommand{\A}{\Sigma}
\newcommand{\size}{\tau}

\newcommand{\lcp}{\mathsf{lcp}}
\newcommand{\lcpzo}{\mathsf{lcp}_{01}}
\newcommand{\lcpz}{\mathsf{lcp}_{0}}
\newcommand{\lcpo}{\mathsf{lcp}_{1}}

\newcommand{\sa}{\mathsf{sa}}
\newcommand{\sazo}{\mathsf{sa}_{01}}
\newcommand{\SA}{\mathsf{sa}}
\newcommand{\LCP}{\mathsf{LCP}}

\newcommand{\XBWT}{\mathsf{xbwt}}
\newcommand{\bwt}{\mathsf{bwt}}

\newcommand{\mbwt}{\mathsf{bwt}}

\newcommand{\bwtzo}{\mathsf{bwt}_{01}}
\newcommand{\onex}{\mathbf{1}}
\newcommand{\zerox}{\mathbf{0}}

\newcommand{\tx}{\mathsf{t}}

\newcommand{\eps}{$\epsilon$}
\def\stri#1{\mbox{\sf #1}}

\newcommand{\tj}[1]{\mathsf{t}_{#1}}

\newcommand{\sj}[1]{\mathsf{s}_{#1}}
\newcommand{\tz}{\mathsf{t}_0}
\newcommand{\tone}{\mathsf{t}_1}
\newcommand{\tzo}{\mathsf{t}_{01}}

\newcommand{\txx}[1]{\mathsf{t}_{#1}}
\newcommand{\sxx}{\mathsf{s}}
\newcommand{\saxx}{\mathsf{sa}}

\newcommand{\nz}{{n_0}}
\newcommand{\none}{{n_1}}
\newcommand{\mz}{{m_0}}
\newcommand{\mone}{{m_1}}
\newcommand{\eos}{\$}
\newcommand{\eosz}{\$_0}
\newcommand{\eosone}{\$_1}

\newcommand{\bwtz}{\mathsf{bwt}_0}
\newcommand{\bwto}{\mathsf{bwt}_1}
\newcommand{\bwtx}[1]{\mathsf{bwt}_{#1}}
\newcommand{\oneb}{{\bf 1}}
\newcommand{\zerob}{{\bf 0}}
\newcommand{\Bid}{\mathsf{Block\_id}}
\newcommand{\bid}{\mathsf{id}}
\newcommand{\avelcp}{\mathsf{aveLcp}_{01}}
\newcommand{\avelcpx}{\mathsf{aveLcp}}
\newcommand{\maxlcpx}{\mathsf{maxLcp}}

\newcommand{\bv}[1]{Z^{(#1)}}

\newcommand{\kh}{{b(h)}}
\newcommand{\sbot}{0}
\newcommand{\Bxx}{B_2}
\newcommand{\zzx}{\mathit{0}}
\newcommand{\oddx}{\mathit{1}}
\newcommand{\evenx}{\mathit{2}}
\newcommand{\oox}{\mathit{3}}

\newcommand{\xbwt}{\mathsf{xbwt}}

\newcommand{\last}{\mathsf{Last}}

\newcommand{\lastz}{\mathsf{Last}_{0}}
\newcommand{\lasto}{\mathsf{Last}_{1}}

\newcommand{\lastj}[1]{\mathsf{Last}_{#1}}

\newcommand{\Lxo}{L_{1}}

\newcommand{\cbwt}{\mathsf{cbwt}}
\newcommand{\cbwtzo}{\cbwt_{01}}
\newcommand{\cbwtz}{\cbwt_{0}}
\newcommand{\cbwto}{\cbwt_{1}}

\newcommand{\csa}{\mathsf{csa}}
\newcommand{\csab}{\csa_b}
\newcommand{\csao}{\csa_1}
\newcommand{\csaz}{\csa_0}
\newcommand{\csazo}{\csa_{01}}
\newcommand{\clcp}{\mathsf{clcp}}
\newcommand{\clcpzo}{\clcp_{01}}

\newcommand{\rotz}[1]{\mathsf{rot}_0(#1)}
\newcommand{\roto}[1]{\mathsf{rot}_1(#1)}
\newcommand{\rotzo}[1]{\mathsf{rot}_{01}(#1)}
\newcommand{\rotb}[1]{\mathsf{rot}_{b}(#1)}
\newcommand{\rotinf}[2]{\mathsf{rot}_{#1}(#2)^\infty}

\newcommand{\hm}{{\sf H\&M}}
\newcommand{\gap}{{\sf Gap}}
\newcommand{\xgap}{{\sf xGap}}
\newcommand{\cgap}{{\sf cGap}}
\newcommand{\cegap}{{\sf \xxy Gap}\xspace}

\newcommand{\avelen}{\mathbf{avelen}}

\newcommand{\useless}{irrelevant}

\def\gSACA{{\sf gSACA-K}}

\newcommand{\purple}{\textcolor{purple}}

\title{Lightweight merging of compressed indices based on {BWT} variants}

\author{Lavinia Egidi}{University of Eastern Piedmont\\{Alessandria, Italy}}{}{}{}
\author{Giovanni Manzini}{University of Eastern Piedmont\\{Alessandria, Italy}}{IIT-CNR\\{Pisa, Italy}}{}{}{}

\authorrunning{L. Egidi, G. Manzini} 

\subjclass{Theory of computation $\rightarrow$ Design and analysis of algorithms}


\EventEditors{John Q. Open and Joan R. Access}
\EventNoEds{2}
\EventLongTitle{42nd Conference on Very Important Topics (CVIT 2016)}
\EventShortTitle{CVIT 2016}
\EventAcronym{CVIT}
\EventYear{2016}
\EventDate{December 24--27, 2016}
\EventLocation{Little Whinging, United Kingdom}
\EventLogo{}
\SeriesVolume{42}
\ArticleNo{23}
\nolinenumbers 
\hideLIPIcs  

\begin{document}

\newtheorem{property}[theorem]{Property}

\maketitle

\begin{abstract}
In this paper we propose a flexible and lightweight technique for merging compressed indices based on variants of Burrows-Wheeler transform (BWT), thus addressing the need for algorithms that compute compressed indices over large collections using a limited amount of working memory. Merge procedures make it possible to use  an incremental strategy for building large indices based on merging indices for progressively larger subcollections. 

Starting with a known lightweight algorithm for merging BWTs  [Holt and McMillan, Bionformatics 2014], we show how to modify it in order to merge, or compute from scratch, also the Longest Common Prefix (LCP) array. We then expand our technique for merging compressed tries and circular/permuterm compressed indices, two compressed data structures for which there were hitherto no known merging algorithms.   
\end{abstract}



\section{Introduction} \label{sec:intro}

The {Burrows Wheeler transform} (BWT), originally introduced as a tool for data compression~\cite{bw}, has found application in the compact representation of many different data structures. After the seminal works~\cite{NM-survey07} showing that the BWT can be used as a compressed full text index for a single string, many researchers have proposed variants of this transformation for string collections~\cite{jda/CoxGRS16,tcs/MantaciRRS07}, trees~\cite{xbw05,jacm09}, graphs~\cite{wabi/BoweOSS12,bioinformatics/MuggliBNMBRGPB17,Siren2017}, and alignments~\cite{Na2015,Na2018}. See~\cite{tcs/GagieMS17} for an attempt to provide a unified view of these variants. 

In this paper we consider the problem of constructing compressed indices for {\em string collections} based on BWT variants. A compressed index is obviously most useful when working with very large amounts of data. Therefore, a fundamental requirement for construction algorithms, in order to be of practical use, is that they are {\em lightweight} in the sense that they use a limited amount of working space, i.e. space in addition to the space used for the input and the output. Indeed, the construction of compressed indices in linear time and small working space is an active and promising area of research, see~\cite{stoc/Belazzougui14,FNNdcc19.3,soda/MunroNN17} and references therein.

A natural approach when working with string collections is to build the indexing data structure {\em incrementally}, that is, for progressively larger subcollections. For example, when additional data should be added to an already large index, the incremental construction appears much more reasonable, and often works better in practice, than rebuilding the complete index from scratch, even when the from-scratch option has better theoretical bounds. Indeed, in~\cite{Siren09} and~\cite{Muggli229641} the authors were able to build the largest indices in their respective fields using the incremental approach.

Along this path, Holt and McMillan~\cite{bioinformatics/HoltM14,bcb/HoltM14} proposed a simple and elegant algorithm, that we call the \hm\ algorithm from now on, for merging BWTs of collections of sequences. For collections of total size $n$,  their fastest version takes $\Oh(n\,\avelcp)$ time where  $\avelcp$ is the average length of the longest common prefix between suffixes in the collection. The average length of the longest common prefix is $\Oh(n)$ in the worst case but $\Oh(\log n)$ for random strings and for many real world datasets~\cite{LMS12}. However, even when $\avelcp = \Oh(\log n)$ the \hm\ algorithm is not theoretically optimal since computing the BWT from scratch takes~$\Oh(n)$ time. Despite its theoretical shortcomings, because of its simplicity and small space usage, the \hm\ algorithm is competitive in practice for collections with relatively small average LCP. In addition, since the \hm\ algorithm accesses all data by sequential scans, it has been adapted to work on very large collections in external memory~\cite{bioinformatics/HoltM14}.

In this paper we revisit the \hm\ algorithm and we show that its main technique can be adapted to solve the merging problem for three different compressed indices based on the BWT. 

First, in Section~\ref{sec:gap} we describe a procedure to merge, in addition to the BWTs, the Longest Common Prefix (LCP) arrays of string collections. The LCP array is often used to provide additional functionalities to indices based on the BWT~\cite{NM-survey07}, and the issue of efficiently computing and storing LCP values has recently received much attention~\cite{GogO13, jea/KarkkainenK16}. Our algorithm has the same $\Oh(n\, \avelcpx)$ complexity as the \hm\ algorithm.  

Next, in Section~\ref{sec:tries} we describe a procedure for merging {\em compressed labelled trees} (tries) as produced by the eXtended BWT transform (XBWT)~\cite{xbw05,jacm09}. This result is particularly interesting since at the moment there are no time and space optimal algorithms for the computation from scratch of the XBWT.  Our algorithm takes time proportional to the number of nodes in the output tree times the {\em average} node height.

Finally, in Section~\ref{zirrs} we describe algorithms for merging {\em compressed indices for circular patterns}~\cite{Honetal11}, and {\em compressed permuterm indices}~\cite{talg/FerraginaV10}. The time complexity of these algorithms is proportional to the total collection size times the average {\em circular LCP}, a notion that naturally extends the LCP to the modified lexicographic order used for circular strings.

Our algorithms are based on the \hm\ technique specialized to the particular features of the different compressed indices given as input. They all make use of techniques to recognize blocks of the input that become irrelevant for the computation and skip them in successive iterations. Because of the skipping of irrelevant blocks we call our merging procedures \gap\ algorithms. Our algorithms are all lightweight in the sense that, in addition to the input and the output, they use only a few bitarrays of working space and the space for handling the irrelevant blocks. The latter amount of space can be significant for pathological inputs, but in practice we found it takes between 2\% and 9\% of the overall space, depending on the alphabet size. 

The \gap\ algorithms share with the \hm\ algorithm the feature of accessing all data by sequential scans and are therefore suitable for implementation in external memory. In~\cite{wabi/EgidiLMT18} an external memory version of the \gap\ algorithm for merging BWT and LCP arrays is engineered, analyzed, and extensively tested on collections of DNA sequences. The results reported there show that the external memory version of \gap\ outperforms the known external memory algorithms for BWT/LCP computation when the avergae LCP of the collection is relatively small or when the strings of the input collection have widely different lengths. 

To the best of our knowledge, the problem of incrementally building compressed indices via merging has been previously addressed only in~\cite{dcc/Siren16} and~\cite{Muggli229641}. Sir{\'{e}}n presents in~\cite{dcc/Siren16} an algorithm that maintains a BWT-based compressed index in RAM and incrementally merges new collections to it. The algorithm is the first that makes it possible to build indices for Terabytes of data without using a specialized machine with a lot of RAM. However, Sir{\'{e}}n's algorithm is specific for a particular compressed index (which doesn't use the LCP array), while ours can be more easily adapted to build different flavors of compressed indices as shown in this paper. 
In~\cite{Muggli229641} the authors present a merge algorithm for colored de Bruijn graphs. Their algorithm is also inspired by the \hm\ algorithm and the authors report a threefold reduction in working space compared to the state of the art methods for from scratch de Bruijn graphs. Inspired by the techniques introduced in this paper, we are currently working on an improved de Bruijn graph merging algorithm~\cite{corr/EgidiLM19} that also supports the construction of succinct {Variable Order} de Bruijn graph representations~\cite{dcc/BoucherBGPS15}.

\ignore{
We propose a flexible technique for the design of lightweight BWT merging algorithms, and instantiate it in three algorithms that extend \hm's (Holt and McMillan's) algorithm to also compute LCP values, to merge XBWTs for tries and to merge cBWTs.

We develop the idea from~\cite[Sect.~2.1]{bcb/HoltM14} of limiting the overall processing time of the sorting procedure by selectively processing at each iteration only regions where the sorting is not yet complete. The impact on the \hm\ algorithm is  a reduction of the asymptotic running time from $\Oh(n\cdot\maxlcpx)$~\cite{bioinformatics/HoltM14} (where $\maxlcpx$ is the maximum length of the longest common prefix) to the stated $\Oh(n\cdot\avelcp)$. We take an approach specular to theirs by focusing on the regions that {\em do not need} further processing. Although this may seem an insignificant difference, it makes all non-sorting operations completely different and allows to fuse {\em \useless} regions which makes the bookkeeping lighter. In addition we introduce a lazy strategy by which the algorithm identifies regions (which we call {\em monochrome}) in which sorting need not be completed because the inputs already provide sufficient information.
 
As a result we show that, applying the aforementioned ideas, the running time of our three algorithms can be reduced from depending on maximal lengths (as in~\cite{bioinformatics/HoltM14}) to being proportional to {\em average} lengths (as in~\cite{bcb/HoltM14}). In particular the algorithm that, in addition to merging the multi-string BWTs, computes the corresponding LCP array (\gap) has the
same asymptotic cost as the variant of \hm\ from~\cite{bcb/HoltM14} and uses additional space only for storing its
additional output, i.e. the LCP values. In our implementation, the \gap\
algorithm uses only $\approx 1.5$ bytes per symbol of workspace in addition
to the input and the output, making it interesting when the overall size of
the collection is close to the available RAM.
The algorithm that merges XBWTs (\xgap) has asymptotic running time proportional to the number of nodes in the output trie times the {\em average} length of the strings in the collections that it represents. \purple{(Si pu\`o dire cos\'i?)} Finally, the algorithm that merges cBWTs (\cgap) works in time proportional to the length of the output cBWT times the {\em average} length of cLCP values (defined as a natural extension of the concept of longest common prefixes). The space bound is for \xgap\ and \cgap\ the same as for the variant of the \hm\ algorithm.

Our algorithms owe their names to the gaps represented by the \useless\ regions.}

\section{Background}\label{sec:notation}

Let $\txx{}[1,n]$ denote a string of length $n$ over an alphabet $\A$ of constant size~$\sigma$. We
write $\txx{}[i,j]$ to denote the substring $\txx{}[i] \txx{}[i+1] \cdots
\txx{}[j]$. If $j\geq n$ we assume $\txx{}[i,j] = \txx{}[i,n]$. If $i>j$ or
$i > n$ then $\txx{}[i,j]$ is the empty string. Given two strings $\txx{}$
and $\sxx$ we write $\txx{} \preceq \sxx$ ($\txx{} \prec \sxx$) to denote
that $\txx{}$ is lexicographically (strictly) smaller than $\sxx$. We denote
by $\LCP(\txx{},\sxx{})$ the length of the longest common prefix between
$\txx{}$ and $\sxx$.

The {\em suffix array} $\saxx[1,n]$ associated to $\txx{}$ is the permutation
of $[1,n]$ giving the lexicographic order of $\txx{}$'s suffixes, that is,
for $i=1,\ldots,n-1$, $\txx{}[\saxx[i],n] \prec \txx{}[\saxx[i+1],n]$. The
{\em longest common prefix} array $\lcp[1,n+1]$ is defined for $i=2,\ldots,n$
by
\begin{equation}\label{eq:lcpdef}
\lcp[i]=\LCP(\txx{}[\saxx[i-1],n],\txx{}[\saxx[i],n]);
\end{equation}
the $\lcp$ array stores the length of the longest common prefix between
lexicographically consecutive suffixes. For convenience we define
$\lcp[1]=\lcp[n+1] = -1$. We also define the maximum and average LCP as:
\begin{equation}
\maxlcpx = \max\nolimits_{1<i\leq n} \lcp[i],\qquad  
\avelcpx = \biggl(\sum\nolimits_{1<i\leq n} \lcp[i]\biggr)/n.
\end{equation}
The {\em Burrows-Wheeler transform} $\bwt[1,n]$ of
$\txx{}$ is defined by
$$
\bwt[i] = \begin{cases}
\tx[n] & \mbox{if } \saxx[i]=1\\
\tx[\saxx[i]-1] & \mbox{if } \saxx[i]>1.
\end{cases}
$$
$\bwt$ is best seen as the permutation of $\txx{}$ in which the position of
$\txx{}[j]$ coincides with the lexicographic rank of $\txx{}[j+1,n]$ (or of
$\txx{}[1,n]$ if $j=n$) in the suffix array. We call the string $\txx{}[j+1,n]$ {\em
context} of $\txx{}[j]$. See Figure~\ref{fig:BWTetc} for an example.

\begin{figure}[t]
\def\xy{$\$_0$}
\def\xz{$\$_1$}
\setlength{\tabcolsep}{5pt}
\begin{center}\sf
\begin{tabular}[t]{ccc}
\begin{tabular}[t]{|r|c|l|}\hline
lcp&bwt& {\em context}\\\hline
 -1 & b & \xy       \\
  0 & c & ab\xy     \\
  2 &\xy& abcab\xy  \\
  0 & a & b\xy      \\
  1 & a & bcab\xy   \\
  0 & b & cab\xy    \\
 -1 &   & \\\hline
\end{tabular}&
\begin{tabular}[t]{|r|c|l|}\hline
lcp&bwt& {\em context}\\\hline
 -1 & c & \xz       \\
  0 &\xz& aabcabc\xz\\
  1 & c & abc\xz    \\
  3 & a & abcabc\xz \\
  0 & a & bc\xz     \\
  2 & a & bcabc\xz  \\
  0 & b & c\xz      \\
  1 & b & cabc\xz   \\
 -1 &   & \\\hline
\end{tabular}&
\begin{tabular}[t]{|r|r|c|l|}\hline
id &$\lcp_{01}$&$\mbwt_{01}$& {\em context}\\\hline
 0 & -1 & b & \xy\\
 1 &  0 & c & \xz       \\
 1 &  0 &\xz& aabcabc\xz\\
 0 &  1 & c & ab\xy\\
 1 &  2 & c & abc\xz    \\
 0 &  3 &\xy& abcab\xy\\
 1 &  5 & a & abcabc\xz \\
 0 &  0 & a & b\xy\\
 1 &  1 & a & bc\xz     \\
 0 &  2 & a & bcab\xy\\
 1 &  4 & a & bcabc\xz  \\
 1 &  0 & b & c\xz      \\
 0 &  1 & b & cab\xy\\
 1 &  3 & b & cabc\xz   \\
   & -1 &   & \\\hline
\end{tabular}
\end{tabular}
\end{center}
\caption{LCP array and BWT for $\txx{0}=\stri{abcab\xy}$
and $\txx{1}=\stri{aabcabc\xz}$,
and multi-string BWT and corresponding LCP array for the
same strings. Column {\sf id} shows, for each entry of
$\mbwt_{01} = \stri{bc\xz cc\xy aaaabbb}$ whether it comes
from $\txx{0}$ or $\txx{1}$.}\label{fig:BWTetc}
\end{figure}


The longest common prefix (LCP) array, and Burrows-Wheeler transform (BWT)
can be generalized to the case of multiple
strings. Historically, the first of such generalizations is the circular BWT~\cite{tcs/MantaciRRS07} considered in Section~\ref{zirrs}. Here we consider the generalization proposed in~\cite{jda/CoxGRS16} which is the one most used in applications. 
Let $\tz[1,\nz]$ and
$\tone[1,\none]$ be such that $\tz[\nz] = \eosz$ and $\tone[\none] = \eosone$
where $\eosz < \eosone$ are two symbols not appearing elsewhere in $\tz$ and
$\tone$ and smaller than any other symbol. Let $\SA_{01}[1,\nz+\none]$ denote
the suffix array of the concatenation $\tz\tone$. The {\em multi-string} BWT
of $\tz$ and $\tone$, denoted by $\mbwt_{01}[1,\nz+\none]$, is defined by
$$
\mbwt_{01}[i] =
\begin{cases}
\tz[\nz]                      & \mbox{if } \SA_{01}[i] = 1\\
\tz[\SA_{01}[i] - 1]       & \mbox{if } 1 < \SA_{01}[i] \leq \nz\\
\tone[\none]                    & \mbox{if } \SA_{01}[i] = \nz + 1\\
\tone[\SA_{01}[i]-\nz - 1] & \mbox{if } \nz +1 < \SA_{01}[i].
\end{cases}
$$
In other words, $\mbwt_{01}[i]$ is the symbol preceding the $i$-th
lexicographically larger suffix, with the exception that if $\SA_{01}[i] = 1$
then $\mbwt_{01}[i] = \eosz$  and if $\SA_{01}[i] = \nz+1$ then
$\mbwt_{01}[i] = \eosone$. Hence, $\mbwt_{01}[i]$ is always a character of the
string ($\tz$ or $\tone$) containing the $i$-th largest suffix (see again
Fig.~\ref{fig:BWTetc}). The above notion of multi-string BWT can be
immediately generalized to define $\mbwt_{1\cdots k}$ for a family of
distinct strings $\txx{1}, \txx{2},\ldots, \txx{k}$. Essentially
$\mbwt_{1\cdots k}$ is a permutation of the symbols in $\txx{1},
\ldots,\txx{k}$ such that the position in $\mbwt_{1\cdots k}$ of $\txx{i}[j]$
is given by the lexicographic rank of its context $\txx{i}[j+1,n_i]$ (or
$\txx{i}[1,n_i]$ if $j=n_i$).

Given the concatenation $\tz\tone$ and its suffix array
$\SA_{01}[1,\nz+\none]$, we consider the corresponding LCP array
$\lcp_{01}[1,\nz+\none+1]$ defined as in~\eqref{eq:lcpdef} (see again
Fig.~\ref{fig:BWTetc}). Note that, for $i=2,\ldots,\nz+\none$, $\lcp_{01}[i]$
gives the length of the longest common prefix between the contexts of
$\mbwt_{01}[i]$ and $\mbwt_{01}[i-1]$. This definition can be immediately
generalized to a family of $k$ strings to define the LCP array
$\lcp_{12\cdots k}$ associated to the multi-string BWT $\mbwt_{12\cdots k}$.

\subsection{The \hm\ Algorithm}\label{sec:hem}

In~\cite{bioinformatics/HoltM14} Holt and McMillan introduced a simple and elegant algorithm, we call it the \hm\ algorithm, to merge multi-string BWTs\footnote{Unless explicitly stated otherwise, in the following we use \hm\ to refer to the algorithm from~\cite{bioinformatics/HoltM14}, and not to its variant proposed in ~\cite{bcb/HoltM14}.}. Because it is the starting point for our results, we now briefly recall its main properties. 

Given $\mbwt_{1\cdots k}$ and $\mbwt_{k+1\,k+2\,\cdots h}$ the \hm\ algorithm computes $\mbwt_{1\cdots h}$. The computation does not explicitly need $\txx{1}, \ldots, \txx{h}$ but only the (multi-string) BWTs to be merged. For
simplicity of notation we describe the algorithm assuming we are merging two
single-string BWTs $\bwtz = \bwt(\tz)$ and $\bwto=\bwt(\tone)$; the same algorithm
works in the general case with multi-string BWTs in input.
Note also that the algorithm can be easily adapted to merge more than two
(multi-string) BWTs at the same time.

\ignore{ Let $\tz$ and $\tone$ denote two strings of length respectively
$\nz=|\tz|$ and $\none = |\tone|$. The \hm\ algorithm computes $\mbwt_{01}$
as defined above given the single string BWTs $\bwtz = \bwt(\tz)$ and
$\bwto=\bwt(\tone)$.

in such a way that the position of $\txx{0}[j]$ (resp. $\txx{0}[j]$) is given
by the lexicographic rank of $\txx{1}[j+1,n_1]$ (resp. $\txx{1}[j+1,n_1]$). }

Computing $\mbwt_{01}$ amounts to sorting the symbols of $\bwtz$ and $\bwto$
according to the lexicographic order of their contexts, where the context of
symbol $\bwtz[i]$ (resp. $\bwto[i]$) is $\tz[\sa_0[i],n_0]$ (resp.
$\tone[\sa_1[i],n_1]$). By construction, the symbols in $\bwtz$ and $\bwto$
are already sorted by context, hence to compute $\mbwt_{01}$ we only need to
merge $\bwtz$ and $\bwto$ without changing the relative order of the symbols
within the two input sequences.

The \hm\ algorithm works in successive iterations. After the $h$-th iteration the entries of $\bwtz$ and $\bwto$ are sorted on the basis of the first $h$ symbols of their context. More formally, the output of the $h$-th iteration is a binary vector $\bv{h}$ containing $n_0=|\tz|$ \zerob's and $n_1 = |\tone|$
\oneb's and such that the following property holds.

\begin{property}\label{prop:hblock}
For $i=1,\ldots, n_0$ and $j=1,\ldots n_1$ the $i$-th \zerob\ precedes the
$j$-th \oneb\ in $\bv{h}$ if and only if
\begin{equation}\label{eq:hblock}
\tz[\sa_0[i], \sa_0[i] + h -1] \;\preceq\; \tone[\sa_1[j], \sa_1[j] + h -1]
\end{equation}
(recall that according to our notation if $\sa_0[i] + h -1>n_0$ then
$\tz[\sa_0[i], \sa_0[i] + h -1]$ coincides with $\tz[\sa_0[i], n_0]$, and
similarly for $\tone$).\qed
\end{property}

Following Property~\ref{prop:hblock} we identify the $i$-th \zerob\ in
$\bv{h}$ with $\bwtz[i]$ and the $j$-th \oneb\ in $\bv{h}$ with $\bwto[j]$ so
that $\bv{h}$ encodes a permutation of $\mbwt_{01}$.
Property~\ref{prop:hblock} is equivalent to stating that we can logically
partition $\bv{h}$ into $\kh+1$ blocks
\begin{equation}\label{eq:Zblocks}
\bv{h}[1,\ell_1],\; \bv{h}[\ell_1+1, \ell_2],\; \ldots,\;
\bv{h}[\ell_\kh+1,n_0+n_1]
\end{equation}
such that each block corresponds to a set of $\mbwt_{01}$ symbols whose
contexts are prefixed by the same length-$h$ string (the symbols with a
context shorter than $h$ are contained in singleton blocks). Within
each block the symbols of $\bwtz$ precede those of $\bwto$, and the context
of any symbol in block $\bv{h}[\ell_j+1, \ell_{j+1}]$ is lexicographically
smaller than the context of any symbol in block $\bv{h}[\ell_k+1,
\ell_{k+1}]$ with $k>j$.

The \hm\ algorithm initially sets $\bv{0} = \zerox^{\nz} \onex^{\none}$:
since the context of every $\mbwt_{01}$ symbol is prefixed by the same
length-0 string (the empty string), there is a single block containing all
$\bwt_{01}$ symbols. At iteration $h$ the algorithm computes $\bv{h+1}$ from
$\bv{h}$ using the procedure in Figure~\ref{fig:HMalgo}. The following lemma is a
restatement of Lemma~3.2 in~\cite{bioinformatics/HoltM14} using our notation (see~\cite{spire/EgidiM17} for a proof in our notation).

\algnewcommand\KwTo{\textbf{to }} \algnewcommand\KwAnd{\textbf{and }}
\begin{figure}
\hrule\smallbreak
\begin{algorithmic}[1]
\State Initialize array $F[1,\sigma]$
\State $k_0 \gets 1$; $k_1 \gets 1$  \Comment{Init counters for $\bwtz$ and $\bwto$}
\For{$k \gets 1$ \KwTo $n_0+ n_1$}
    \State $b \gets \bv{h-1}[k]$\Comment{Read bit $b$ from $\bv{h-1}$}
    \State $c \gets \bwt_b[k_b{\mathsf ++}]$ \Comment{Get symbol from $\bwtz$ or $\bwto$ according to $b$}
    \If {$c \neq \eos$} 
      \State $j \gets F[c]{\mathsf ++}$ \Comment{Get destination for $b$ according to symbol $c$}      \label{line:movenoeos}
    \Else
      \State $j \gets b$  \Comment{Symbol $\eos_b$ goes to position~$b$}\label{line:moveeos}
    \EndIf
    \State $\bv{h}[j] \gets b$ \Comment{Copy bit $b$ to $\bv{h}$}
\EndFor
\end{algorithmic}
\smallbreak\hrule
\caption{Main loop of algorithm \hm\ for computing $\bv{h}$ given
$\bv{h-1}$. Array $F$ is initialized so that $F[c]$ contains
the number of occurrences
of symbols smaller than $c$ in $\bwtz$ and $\bwto$ plus one. Note that 
the bits stored in $\bv{h}$ immediately after reading symbol $c\neq\eos$
are stored in positions from $F[c]$ to $F[c+1]-1$ of $\bv{h}$.}\label{fig:HMalgo}
\end{figure}

\begin{lemma}\label{lemma:hblock}
For $h=0,1,2,\ldots$ the bit vector $\bv{h}$ satisfies
Property~\ref{prop:hblock}.\qed
\end{lemma}

\ignore{
\begin{proof}
We prove the
result by induction. For $h=0$ and $\delta=0,1$,
$\txx{\delta}[\sa_\delta[i],\sa_\delta[i]-1]$ is the empty string
so~\eqref{eq:hblock} is always true and Property~\ref{prop:hblock} is
satisfied by $\bv{0} = \zerox^\nz \onex^\none$.

To prove the ``if'' part, let $h>0$ and let $1 \leq v < w \leq \nz+\none$
denote two indexes such that $\bv{h}[v]$ is the $i$-th \zerob\ and
$\bv{h}[w]$ is the $j$-th \oneb\ in $\bv{h}$. We need to show that under
these assumptions inequality~\eqref{eq:hblock} on the lexicographic order
holds.
Assume first $\tz[\sa_0[i]]\neq \tone[\sa_1[j]]$. The hypothesis $v<w$
implies $\tz[\sa_0[i]]< \tone[\sa_1[j]]$ hence~\eqref{eq:hblock} certainly
holds.

Assume now $\tz[\sa_0[i]] = \tone[\sa_1[j]]$. We preliminarily observe that
it must be $\sa_0[i]\neq\nz$ and $\sa_1[i]\neq\none$: otherwise we would have
$\tz[\sa_0[i]]=\xx_0$ or $\tone[\sa_1[j]]=\xx_1$ which is impossible since
these symbols appear only once in $\tz$ and $\tone$.

Let $v'$, $w'$ denote respectively the value of the main loop variable~$k$ in
the procedure of Figure~\ref{fig:HMalgo} when the entries $\bv{h}[v]$ and
$\bv{h}[w]$ are written (hence, during the scanning of $\bv{h-1}$). The
hypothesis $v<w$ implies ${v}' < {w}'$. By construction
$\bv{h-1}[{v}']=\zerox$ and $\bv{h-1}[{w}']=\onex$. Say ${v}'$ is the $i'$-th
\zerob\ in $\bv{h-1}$ and ${w}'$ is the $j'$-th \oneb\ in $\bv{h-1}$. By the
inductive hypothesis on $\bv{h-1}$ we have
\begin{equation}\label{eq:hblock2}
\tz[\sa_0[i'], \sa_0[i'] + h -2] \;\preceq\; \tone[\sa_1[j'], \sa_1[j'] + h -2],
\end{equation}
The fundamental observation is that, being $\sa_0[i]\neq\nz$ and
$\sa_1[j]\neq\none$, it is
$$
\sa_0[i'] = \sa_0[i] + 1\qquad\mbox{and}\qquad
\sa_1[j'] = \sa_1[j] + 1.
$$
Since
\begin{align}
\tz[\sa_0[i],\sa_0[i]+h-1]\; &= \;\tz[\sa_0[i]]\tz[\;\sa_0[i'],\sa_0[i']+h-2]\\
\tone[\sa_1[j],\sa_1[j]+h-1]\; &=\; \tone[\sa_1[j]]\;\tone[\sa_1[j'],\sa_1[j']+h-2]
\end{align}
combining $\tz[\sa_0[i]] = \tone[\sa_1[j]]$ with~\eqref{eq:hblock2} gives
us~\eqref{eq:hblock}.

For the ``only if'' part assume~\eqref{eq:hblock} holds. We need to prove
that in $\bv{h}$ the $i$-th \zerob\ precedes the $j$-th \oneb. If
$\tz[\sa_0[i]] < \tone[\sa_1[j]]$ the proof is immediate. If $\tz[\sa_0[i]] =
\tone[\sa_1[j]]$, we must have
$$
\tz[\sa_0[i]+1,\sa_0[i]+h-1] \preceq
\tone[\sa_1[j]+1,\sa_1[j]+h-1].
$$
By induction, if $\sa_0[i'] = \sa_0[i] + 1$ and $\sa_1[j']= \sa_1[j]+1$ in
$\bv{h-1}$ the $i'$-th \zerob\ precedes the $j'$-th \oneb. During iteration~$h$,
the $i$-th \zerob\ in $\bv{h}$ is written when processing the $i'$-th \zerob\
of $\bv{h-1}$, and the $j$-th \oneb\ in $\bv{h}$ is written when processing
the $j'$-th \oneb\ of $\bv{h-1}$. Since in $\bv{h-1}$ the $i'$-th \zerob\
precedes the $j'$-th \oneb\ and
$$
\bwtz[i']=\tz[\sa_0[i]] = \tone[\sa_1[j]] = \bwto[j']
$$
in $\bv{h}$ their relative order does not change and the $i$-th \zerob\
precedes the $j$-th \oneb\ as claimed.\qed
\end{proof}
}

\begin{figure}
\hrule\smallbreak
\begin{algorithmic}[1]
\State Initialize arrays $F[1,\sigma]$ and $\Bid[1,\sigma]$\label{line:init}
\State $k_0 \gets 1$; $k_1 \gets 1$  \Comment{Init counters for $\bwtz$ and $\bwto$}
\For{$k \gets 1$ \KwTo $n_0+ n_1$}
    \If{$B[k]\neq \sbot $ \KwAnd $B[k]\neq h$}\label{line:B=0h}
      \State $\bid\gets k$\Comment{A new block of $\bv{h-1}$ is starting}\label{line:blockstart}
    \EndIf\label{line:B=0hend}
    \State $b \gets \bv{h-1}[k]$\Comment{Read bit $b$ from $\bv{h-1}$}\label{line:block_process_start}
    \State $c \gets \bwt_b[k_b{\mathsf ++}]$ \Comment{Get symbol from $\bwtz$ or $\bwto$ according to $b$}
    \If {$c \neq \eos$} \label{line:HMlcp:noteos}
      \State $j \gets F[c]{\mathsf ++}$ \Comment{Get destination for $b$ according to symbol $c$}
    \Else
      \State $j \gets b$  \Comment{Symbol $\eos_b$ goes to position~$b$} \label{line:HMlcp:eos}
    \EndIf
    \State $\bv{h}[j] \gets b$ \Comment{Copy bit $b$ to $\bv{h}$}\label{line:hem:updatebv}
    \If{$\Bid[c]\neq \bid$} \label{line:bid_start}
      \State $\Bid[c]\gets \bid$\Comment{Update block id for symbol $c$}
      \If{$B[j] = \sbot$} \Comment{Check if already marked}\label{line:new_start}
        \State$B[j] = h$\Comment{A new block of $\bv{h}$ will start here}\label{line:writeh}
      \EndIf
    \EndIf \label{line:block_process_end}
\EndFor
\end{algorithmic}
\smallbreak\hrule
\caption{Main loop of the \hm\ algorithm modified for the computation of
the $\lcp$ values. At Line~\ref{line:init}
for each symbol $c$ we set $\Bid[c] = -1$ and $F[c]$ as in
Figure~\ref{fig:HMalgo}. At the beginning of the algorithm we initialize the
array $B[1,\nz+\none+1]$ as $B = 1\:0^{\nz+\none-1}\:1$.}\label{fig:HMlcp}
\end{figure}

\section{Computing LCP values with the \hm\ algorithm}\label{sec:hmlcp}

Our first result is to show that with a simple modification to the \hm\ algorithm it is possible to  
compute the LCP array~$\lcpzo$, in addition to merging $\bwtz$ and $\bwto$.
Our strategy consists in
keeping explicit track of the logical blocks we have defined for $\bv{h}$ and represented in~\eqref{eq:Zblocks}. We maintain an integer array
$B[1,n_0+n_1+1]$ such that at the end of iteration $h$ it is $B[i]\neq 0$ if and only if a block of $\bv{h}$ starts at position~$i$. The use of such integer array is shown in Figure~\ref{fig:HMlcp}. Note that: $(i)$ initially we set $B = 1\:0^{\nz+\none-1}\:1$ and once an entry in $B$ becomes nonzero it is never changed,  $(ii)$ during iteration $h$ we only write to $B$ the value $h$,
$(iii)$ because of the test at Line~\ref{line:B=0h} the values written during
iteration $h$ influence the algorithm only in subsequent iterations. In order to identify new blocks, we maintain
an array $\Bid[1,\sigma]$ such that $\Bid[c]$ is the $\bid$ of the block of
$\bv{h-1}$ to which the last seen occurrence of symbol $c$ belonged.

The following lemma shows that the nonzero values of $B$ at the end of
iteration~$h$ mark the boundaries of $\bv{h}$'s logical blocks.


\begin{lemma} \label{lemma:B}
For any $h\geq 0$, let $\ell$, $m$ be such that $1 \leq \ell \leq m \leq
\nz+\none$ and
\begin{equation}\label{eq:lcpblock}
\lcpzo[\ell] < h,\quad \min(\lcpzo[\ell+1], \ldots, \lcpzo[m]) \geq h,
\quad \lcpzo[m+1] < h.
\end{equation}
Then, at the end of iteration $h$ the array $B$ is such that
\begin{equation}\label{eq:Bblock}
B[\ell]\neq 0, \quad B[\ell+1] = \cdots = B[m] = 0,
\quad B[m+1] \neq 0
\end{equation}
and $\bv{h}[\ell,m]$ is one of the blocks in~\eqref{eq:Zblocks}.
\end{lemma}

\begin{proof}
We prove the result by induction on $h$. For $h=0$, hence before the
execution of the first iteration, \eqref{eq:lcpblock} is only valid for $\ell=1$
and $m=\nz+\none$ (recall that we defined $\lcpzo[1]=\lcpzo[\nz+\none+1]=-1$).
Since initially $B=1\: 0^{\nz+\none-1}\:1$ our claim holds.

Suppose now that~\eqref{eq:lcpblock} holds for some $h>0$. Let
$s=\tzo[\sazo[\ell],\sazo[\ell] + h-1]$; by~\eqref{eq:lcpblock} $s$ is a
common prefix of the suffixes starting at positions $\sazo[\ell]$,
$\sazo[\ell+1]$, \ldots, $\sazo[m]$, and no other suffix of $\tzo$ is
prefixed by~$s$. By Property~\ref{prop:hblock} the \zerob s and \oneb s in
$\bv{h}[\ell,m]$ corresponds to the same set of suffixes That is, if $\ell
\leq v \leq m$ and $\bv{h}[v]$ is the $i$th \zerob\ (resp. $j$th \oneb) of
$\bv{h}$ then the suffix starting at $\tz[\sa_0[i]]$ (resp.
$\tone[\sa_1[j]]$) is prefixed by $s$.

To prove~\eqref{eq:Bblock} we start by showing that, if $\ell < m$, then at
the end of iteration $h-1$ it is $B[\ell+1] = \cdots = B[m] = 0$. To see this
observe that the range $\sazo[\ell,m]$ is part of a (possibly) larger range
$\sazo[\ell',m']$ containing all suffixes prefixed by the length $h-1$ prefix
of $s$. By inductive hypothesis, at the end of iteration $h-1$ it is $B[\ell'+1]
= \cdots = B[m'] = 0$ which proves our claim since $\ell'\leq \ell$ and $m
\leq m'$.

To complete the proof, we need to show that during iteration $h$: $(i)$ we do not
modify $B[\ell+1,m]$ and $(ii)$ we write a nonzero to
$B[\ell]$ and $B[m+1]$ if they do not already contain a nonzero. Let $c=s[0]$
and $s'=s[1,h-1]$ so that $s = cs'$. Consider now the range $\sazo[e,f]$
containing the suffixes prefixed by $s'$. By inductive hypothesis at the end
of iteration $h-1$ it is
\begin{equation}\label{eq:Bblock_inproof}
B[e]\neq 0, \quad B[e+1] = \cdots = B[f] = 0,
\quad B[f+1] \neq 0.
\end{equation}
During iteration $h$, the bits in $\bv{h}[\ell,m]$ are possibly changed only
when we are scanning the region $\bv{h-1}[e,f]$ and we find an entry
$b=\bv{h-1}[k]$, $e\leq k \leq f$, such that the corresponding value in
$\bwtx{b}$ is $c$. Note that by~\eqref{eq:Bblock_inproof} as soon as $k$
reaches $e$ the variable $\bid$ changes and becomes different from all values
stored in $\Bid$. Hence, at the first occurrence of symbol $c$ the value $h$
will be stored in $B[\ell]$ (Line~\ref{line:writeh}) unless a nonzero is
already there. Again, because of~\eqref{eq:Bblock_inproof}, during the
scanning of $\bv{h-1}[e,f]$ the variable $\bid$ does not change so subsequent
occurrences of $c$ will not cause a nonzero value to be written to
$B[\ell+1,m]$. Finally, as soon as we leave region $\bv{h-1}[e,f]$ and $k$
reaches $f+1$, the variable $\bid$ changes again and at the next occurrence
of $c$ a nonzero value will be stored in $B[m+1]$. If there are no more
occurrences of $c$ after we leave region $\bv{h-1}[e,f]$ then either
$\sazo[m+1]$ is the first suffix array entry prefixed by symbol $c+1$ or
$m+1=\nz+\none+1$. In the former case $B[m+1]$ gets a nonzero value at iteration 1, in the latter case $B[m+1]$ gets a nonzero value when we initialize array $B$. \qed
\end{proof}

\begin{corollary}\label{cor:lcp}

For $i=2,\ldots,\nz+\none$, if $\lcpzo[i] = \ell$, then starting from the end
of iteration $\ell+1$ it is $B[i]=\ell+1$.

\end{corollary}

\begin{proof}
By Lemma~\ref{lemma:B} we know that $B[i]$ becomes nonzero only after iteration
$\ell+1$. Since at the end of iteration $\ell$ it is still $B[i]=0$ during iteration
$\ell+1$ $B[i]$ gets the value $\ell+1$ which is never changed in successive
iterations.\qed
\end{proof}

The above corollary suggests the following algorithm to compute $\bwtzo$ and $\lcpzo$: repeat the procedure of Figure~\ref{fig:HMlcp} until the iteration $h$
in which all entries in $B$ become nonzero. At that point $\bv{h}$ describes
how $\bwtz$ and $\bwto$ should be merged to get $\bwtzo$ and for $i=2,\ldots,
\nz+\none$ $\lcpzo[i] = B[i] - 1$. The above strategy requires a number of iterations, each one taking $\Oh(\nz+\none)$ time, equal to the maximum of the $\lcp$ values, for an overall complexity of $\Oh((\nz+\none)\,\mathsf{maxlcp}_{01})$, where $\mathsf{maxlcp}_{01}=\max_i \lcpzo[i]$. Note that in addition to the space for the input and the output the algorithm only uses two bit arrays (one for the current and the next $\bv{\cdot}$) and a constant number of counters (the arrays $F$ and $\Bid$). Summing up we have the following result. 

\begin{lemma} \label{lemma:hmlcp}
Given $\bwtz$ and $\bwto$, 
the algorithm in Figure~\ref{fig:HMlcp} computes $\bwtzo$
and $\lcpzo$ in $\Oh(n\, \maxlcpx )$ time and $2n + \Oh(\log n)$ bits of working space, where $n = |\tzo|$ and $\maxlcpx=\max_i \lcpzo[i]$ is the maximum LCP of~$\tzo$.\qed
\end{lemma}


\ignore{In the next section we describe a much faster algorithm that
avoids processing the portions of $B$ and $\bv{h}$ which are no longer relevant for the computation of the final result.}

\section{The \gap\ BWT/LCP merging Algorithm}\label{sec:gap}

The \gap\ algorithm, as well as its variants described in the following sections, are based on the notion of {\em monochrome blocks}.

\begin{definition}
If $B[\ell]\neq 0$, $B[m+1]\neq 0$ and $B[\ell+1] = \cdots = B[m] = 0$, we
say that block $\bv{h}[\ell,m]$ is {\em monochrome} if it contains only
\zerob's or only \oneb's.\qed
\end{definition}

Since a monochrome block only contains suffixes from either $\tz$ or $\tone$,
whose relative order is 
known, it does not need to be further
modified.  If in addition, the LCP arrays of $\tz$ and $\tone$ are given in input, then also LCP values inside monochrome blocks are known without further processing.
This intuition is formalized by the following lemmas.

\begin{lemma}\label{lemma:monochrome}
If at the end of iteration $h$ bit vector $\bv{h}$ contains only monochrome
blocks we can compute $\bwtzo$ and $\lcpzo$ in $\Oh(\nz + \none)$ time from $\bwtz$, $\bwto$, $\lcpz$ and $\lcpo$.
\end{lemma}

\begin{proof}
By Property~\ref{prop:hblock}, if we identify the $i$-th \zerob\ in $\bv{h}$
with $\bwtz[i]$ and the $j$-th \oneb\ with $\bwto[j]$ the only elements which
could be not correctly sorted by context are those within the same block.
However, if the blocks are monochrome all elements belong to either $\bwtz$
or $\bwto$ so their relative order is correct.

To compute $\lcpzo$ we observe that if $B[i]\neq 0$ then by (the proof of)
Corollary~\ref{cor:lcp} it is $\lcpzo[i] = B[i] -1$. If instead $B[i]=0$ we
are inside a block hence $\sazo[i-1]$ and $\sazo[i]$ belong to the same
string $\tz$ or $\tone$ and their LCP is directly available in $\lcpz$ or
$\lcpo$.\qed 
\end{proof}

Notice that a lazy strategy of not completely processing monochrome blocks, makes it impossible to compute LCP values from scratch. In this case, in order to compute $\lcpzo$ it is necessary that the algorithm also takes $\lcpo$ and $\lcpz$ in input.

\begin{lemma}\label{lemma:skip}
Suppose that, at the end of iteration $h$, $\bv{h}[\ell,m]$ is a monochrome
block. Then $(i)$ for $g>h$, $\bv{g}[\ell,m] = \bv{h}[\ell,m]$, and $(ii)$
processing $\bv{h}[\ell,m]$ during iteration $h+1$ creates a set of monochrome
blocks in $\bv{h+1}$.
\end{lemma}


\begin{proof}
The first part of the Lemma follows from the observation that subsequent
iterations of the algorithm will only reorder the values within a block (and
possibly create new sub-blocks); but if a block is monochrome the reordering
will not change its actual content.

For the second part, we observe that during iteration $h+1$ as $k$ goes from
$\ell$ to $m$ the algorithm writes to $\bv{h+1}$ the same value which is in
$\bv{h}[\ell,m]$. Hence, a new monochrome block will be created for each
distinct symbol encountered (in $\bwtz$ or $\bwto$) as $k$ goes through the
range $[\ell,m]$.\qed
\end{proof}

The lemma implies that, if block $\bv{h}[\ell,m]$ is monochrome at the end of
iteration $h$, starting from iteration $g=h+2$ processing the range $[\ell,m]$ will
not change $\bv{g}$ with respect to $\bv{g-1}$. Indeed, by the lemma the
monochrome blocks created in iteration $h+1$  do not change in subsequent iterations
(in a subsequent iteration a monochrome block can be split in sub-blocks, but the
actual content of the bit vector does not change). The above observation
suggests that, after we have processed block $\bv{h+1}[\ell,m]$ in iteration
$h+1$, we can mark it as {\em \useless} and avoid to process it again. As the
computation goes on, more and more blocks become \useless. Hence, at the
generic iteration $h$ instead of processing the whole $\bv{h-1}$ we process only the blocks which are still ``active'' and skip \useless\ blocks. Adjacent
\useless\ blocks are merged so that among two active blocks there is at most
one \useless\ block (the {\em gap} after which the algorithm is named). The
overall structure of a single iteration is shown in Figure~\ref{fig:gap2}. The
algorithm terminates when there are no more active blocks since this implies
that all blocks have become monochrome and by Lemma~\ref{lemma:monochrome} we
are able to compute $\bwtzo$ and $\lcpzo$.

\ignore{As we already pointed out in the Introduction, \gap\  keeps explicit track of the \useless\ blocks while \hm\ keeps explicit track of the active blocks (called buckets in~\cite{bcb/HoltM14}). So, as we have seen, \useless\ blocks can be merged creating in $\bv{h}$ wider and wider gaps that need not be processed.}

\begin{figure}
\hrule\smallbreak
\begin{algorithmic}[1]
\If{(next block is \useless)}
    \State{skip it}\label{line:skip}
\Else
    \State{process block}\label{line:process_block}
    \If{(processed block is monochrome)}
      \State mark it \useless
    \EndIf
\EndIf
\If{(last two blocks are \useless)}
    \State{merge them}
\EndIf
\end{algorithmic}
\smallbreak\hrule
\caption{Main loop of the \gap\ algorithm. The processing of active
blocks at Line~\ref{line:process_block} is done as in
Lines~\ref{line:block_process_start}--\ref{line:block_process_end}
of Figure~\ref{fig:HMlcp}.}\label{fig:gap2}
\end{figure}

We point out that at Line~\ref{line:skip} of the \gap\ algorithm we cannot
simply skip an \useless\ block ignoring its content. To keep the algorithm
consistent we must correctly update the global variables of the main loop,
i.e. the array $F$ and the pointers $k_0$ and $k_1$ in
Figure~\ref{fig:HMlcp}. To this end a simple approach is to store for each
\useless\ block the number of occurrences $o_c$ of each symbol $c\in\A$ in
it and the pair $(r_0,r_1)$ providing the number of \zerob's and \oneb's in
the block (recall that an \useless\ block may consist of adjacent monochrome
blocks coming from different strings). When the algorithm reaches an
\useless\ block, $F$, $k_0$, $k_1$ are updated setting $k_0 \gets k_0 + r_0$,
$k_1 \gets k_1 + r_1$ and $\forall c$ $F[c] \gets F[c] + o_c$. The above scheme for handling \useless\ blocks is simple and effective for most applications. However, for a large non-constant alphabet it would imply a multiplicative $\Oh(\sigma)$ slowdown. In~\cite[Sect.~4]{spire/EgidiM17} we present a different scheme for large alphabets with a slowdown reduced to~$\Oh(\log\sigma)$.

We point out that our \gap\ algorithm is related to the \hm\ variant with $\Oh(n\, \avelcpx )$ time complexity described in~\cite[Sect.~2.1]{bcb/HoltM14}: Indeed, the sorting operations are essentially the same in the two algorithms. The main difference is that \gap\ keeps explicit track of the \useless\ blocks while \hm\ keeps explicit
track of the active blocks (called buckets in~\cite{bcb/HoltM14}): this difference makes the non-sorting operations completely different. An advantage of working with \useless\ blocks is that they can be easily merged,
while this is not the case for the active blocks in \hm. Of course, the main difference is that \gap\ merges simultaneously BWT {\em and} LCP values.

\begin{theorem}\label{theo:gap}
Given $\bwtz, \lcpz$ and $\bwto, \lcpo$ let $n=|\bwtz|+\bwto|$.
The \gap\ algorithm computes $\bwtzo$ and $\lcpzo$ in $\Oh(n\, \avelcp )$ time, where $\avelcp=(\sum_i \lcpzo[i])/n$ is the average LCP of the string $\tzo$. The working space is $2n + \Oh(\log n)$ bits, plus the space used for handling irrelevant blocks.
\end{theorem}

\begin{proof}
For the running time we reason as in~\cite{bcb/HoltM14} and observe that the sum, over all iterations, of the length of all active blocks is bounded by $\Oh(\sum_i \lcpzo[i]) = \Oh(n\, \avelcp)$. The time bound follows observing that at any iteration the cost of processing an active block of length $\ell$ is bounded by $\Oh(\ell)$ time.

For the analysis of the working space we observe for the array $B$ we can use the space for the output LCP, hence the working space consists only in $2n$ bits for two instances of the arrays $\bv{\cdot}$ and a constant number of counters (the arrays $F$ and $\Bid$).\qed
\end{proof}

\ignore{
If we are simultaneously merging $k$ BWTs, the only change in the algorithm
is that the arrays $\bv{h}$ must now store integers in $[1,k]$. As a consequence, the overall
running time is still $\Oh( n \log(\sigma) \avelcpx)$ where $n=\sum_i n_i$ is
the size of the merged BWT and $\avelcpx$ is the average of the values in the
merged LCP array. 
}

It is unfortunately impossible to give a clean bound for the space needed for keeping track of irrelevant blocks. Our scheme uses $\Oh(1)$ words per block, but in the worst case we can have $\Theta(n)$ blocks. Although such worst case is rather unlikely, it is important to have some form of control on this additional space. We use the following simple heuristic: we choose a threshold~$\size$ and we keep track of an irrelevant block only if its size is at least~$\size$. This strategy introduces a $\Oh(\size)$ time slowdown but ensures that there are at most $n/(\size+1)$ irrelevant blocks simultaneously. The experiments in the next section show that in practice the space used to keep track of irrelevant blocks is less than 10\% of the total.

\ignore{In the next section we experimentally measure the influence of $\size$ on the space and running time of the algorithm. For our experiments we set $\size \geq \sigma + k$ with the rationale that in our implementation skipping a block requires $\Oh(\sigma + k)$ time, so there is no advantage in skipping a block smaller than $\sigma + k$. We show that in practice the space used to keep track of irrelevant blocks is less than 10\% of the total.}

Note that also in~\cite{bcb/HoltM14} the authors faced the problem of
limiting the memory used to keep track of the active blocks. They suggested
the heuristic of keeping track of active blocks only after the $h$-th
iteration ($h=20$ for their dataset).

\def\ERAH{{\sf Illumina}}
\def\PBclr{{\sf Pacbio}}
\def\Prots{{\sf Proteins}}
\def\Wikit{{\sf Wiki-it}}

\begin{table}[t]
\begin{center}\setlength{\tabcolsep}{7pt}
\begin{tabularx}{\textwidth}{|X|r|r|r|r|r|r|}
\hline
{\sf Name}&{\sf Size GB}&$\sigma$& {\sf Max Len}&{\sf Ave Len} & {\sf Max LCP} & {\sf Ave LCP}\\\hline
\PBclr         &6.24 & 5  & 40212&9567.43& 1055& 17.99\\
\ERAH          &7.60 & 6  &   103& 102.00&  102& 27.53\\
\Wikit         &4.01 & 210&553975&4302.84&93537& 61.02\\
\Prots         &6.11 & 26 & 35991& 410.22&25065&100.60\\
\hline
\end{tabularx}
\end{center}\caption{Collections used in our experiments sorted by average LCP.
Columns 4 and 5 refer to the lengths of the single documents. \PBclr\ are NGS
reads from a {\it D.melanogaster} dataset. \ERAH\ are NGS reads from Human
ERA015743 dataset. \Wikit\ are pages from Italian Wikipedia. \Prots\ are
protein sequences from Uniprot. Collections and source files are available on
\url{https://people.unipmn.it/manzini/gap}.\label{tab:files}}

\begin{center}\setlength{\tabcolsep}{7pt}
\begin{tabularx}{\textwidth}{|X|r|c|rr|rr|rr|}
\hline {\sf Name}             &$k$ &\multicolumn{1}{c|}{\gSACA}
                           &\multicolumn{2}{c|}{$\size=50$}
                           &\multicolumn{2}{c|}{$\size=100$}
                           &\multicolumn{2}{c|}{$\size=200$}\\
               &    & {+$\Phi$}
                           & time&space& time&space& time&space \\\hline
\PBclr         & 7  & 0.46 & 0.41& 4.35& 0.46&4.18 & 0.51&4.09 \\
\ERAH          & 4  & 0.48 & 0.93& 3.31& 1.02&3.16 & 1.09&3.08 \\
\Wikit         & 5  & 0.41 &  ---&  ---&  ---& --- & 3.07&6.55 \\
\Prots         & 4  & 0.59 & 3.90& 4.55& 5.18&4.29 & 7.05&4.15 \\
\hline
\end{tabularx}
\end{center}\caption{For each collection we report the number $k$ of
subcollections, the average running time of \gSACA+$\Phi$ in $\mu$secs per
symbol, and the running time ($\mu$secs) and space usage (bytes) per symbol
for \gap\ for different values of the $\size$ parameter. All tests were executed on a desktop with 32GB RAM and eight Intel-I7 3.40GHz CPUs, using a single CPU in each experiment.\label{tab:times}}
\end{table}

\subsection{Experimental Results}

We have implemented the \gap\ algorithm in C and tested it on the collections shown in Table~\ref{tab:files} which have documents of different size, LCP, and alphabet size. We represented LCP values with the minimum possible number of bytes for each collection: 1 byte for \ERAH, 2 bytes for \PBclr\ and \Prots, and 4 bytes
for \Wikit. We always used 1 byte for each BWT value and $n$ bytes to
represent a pair of $\bv{h}$ arrays using 4 bits for each entry so that the tested implementation can merge simultaneously up to $16$ BWTs. 

Referring to Table~\ref{tab:times}, we split each collection into $k$
subcollections of size less than 2GB and we computed the multi-string SA of
each subcollection using \gSACA~\cite{dcc/LouzaGT16}. From the SA we computed
the multi-string BWT and LCP arrays using the $\Phi$
algorithm~\cite{lcp_cpm09} (implemented in \gSACA). This computation used 13
bytes per input symbol. Then, we merged the subcollections BWTs
and LCPs using \gap\ with different values of the parameter $\size$ which
determines the size of the smallest irrelevant block we keep track of. Since skipping a block takes time proportional to $\sigma+k$, regardeless of~$\size$ \gap\ never keeps track of blocks smaller than that threshold; therefore for \Wikit\ we performed a single experiment where the smallest irrelevant block size was $\sigma+k = 215$.


From the results in Table~\ref{tab:times} we see that \gap's running time is
indeed roughly proportional to the average LCP. For example, \PBclr\ and
\ERAH\ collections both consist of DNA reads but, despite \PBclr\ reads being
longer and having a larger maximum LCP, \gap\ is twice as fast on them
because of the smaller average LCP. Similarly, \gap\ is faster on \Wikit\
than on \Prots\ despite the latter collection having a smaller alphabet and
shorter documents.

\ignore{\gSACA\ running time is not significantly influenced by
the average LCP. If we compare \gap\ with \gSACA\ we see that only in one
instance, \PBclr\ with $\size=50$, \gap\ is faster than \gSACA\ in terms of
$\mu$secs per input symbol. However, since \gap\ is designed to post-process
\gSACA\ output, the comparison of the running time is only important to the
extent that \gap\ is not a bottleneck in our two-step strategy to compute the
multi-string BWT and LCP arrays: the experiments show that this is not the case.
We point out that on our 32GB machine, \gSACA\ cannot compute the
multi-string SA for any of the collections since for inputs larger that 2GB
it uses 9 bytes per input symbol.}

As expected, the parameter $\size$ offers a time-space tradeoff for the \gap\
algorithm. In the space reported in Table~\ref{tab:times}, the fractional
part is the peak space usage for \useless\ blocks, while the integral value is
the space used by the arrays $\bwtx{i}$, $B$ and $\bv{h}$. For example, for
\Wikit\ we use $n$ bytes for the BWTs, $4n$ bytes for the LCP values (the $B$
array), $n$ bytes for $\bv{h}$, and the remaining $0.55n$ bytes are mainly
used for keeping track of \useless\ blocks. This is a relatively high value, about 9\% of the total space,  since in our current implementation the storage of a block grows linearly with the alphabet size. For DNA sequences and $\size=200$ the cost of storing blocks is less than 3\% of the total without a significant slowdown in the running time.

For completeness, we tested the \hm\ implementation from~\cite{bcb/HoltM14}
on the \PBclr\ collection. The running time was 14.57 $\mu$secs per symbol
and the space usage $2.28$ bytes per symbol. These values are only partially
significant for several reasons: $(i)$ \hm\ computes the BWT from scratch,
hence doing also the work of \gSACA, $(ii)$ \hm\ doesn't compute the LCP
array, hence the lower space usage, $(iii)$ the algorithm is implemented in
Cython which makes it easier to use in a Python environment but is not as
fast and space efficient as C.

\subsection{Merging only BWTs}\label{sec:B2}

If we are not interested in LCP values but we only need to merge BWTs, we can still use \gap\ instead of \hm\ to do the computation in $\Oh( n \,\avelcpx)$ time. In that case however, the use of the integer array $B$ recording LCP values is wasteful. We can save space replacing it with an array $\Bxx[1,\nz + \none+1]$ containing two bits per entry representing four possible states called $\{\zzx,\oddx,\evenx,\oox\}$. The rationale for this is that, if we are not interested in LCP values, the entries of $B$ are only used in Line~\ref{line:B=0h} of Fig.~\ref{fig:HMlcp}  where it is tested whether they are different from 0 or $h$. 

During iteration $h$, the values in $\Bxx$ are used instead of the ones in $B$ as follows: An entry $\Bxx[i]=\zzx$ corresponds to $B[i]=0$, an entry $\Bxx[i]=\oox$ corresponds to $0 < B[i] < h-1$. If $h$ is even, an entry $\Bxx[i]=\evenx$ corresponds to $B[i]=h$ and an entry $\Bxx[i]=\oddx$ corresponds to $B[i]=h-1$; while if $h$ is odd the correspondence is $\evenx \rightarrow h-1$, $\oddx \rightarrow h$.  The array $\Bxx$ is initialized as $\oox (\zzx)^{\nz+\none-1}(\oox)$, and it is updated appropriately  in lines~\ref{line:new_startB2}--\ref{line:writehB2}. The reason for this apparently involved scheme is that during iteration $h$, an entry in $\Bxx$ can be modified either before or after we read it at Line~\ref{line:B=0h}. The resulting code is shown in Fig.~\ref{fig:HMBxx}. Using the array $\Bxx$ we can still define (and skip) monochrome blocks and therefore achieve the  $\Oh( n \, \avelcpx)$ complexity.

\begin{figure}
\hrule\smallbreak
\begin{algorithmic}[1]
\setcounter{ALG@line}{3}
    \If{$\Bxx[k]\neq \zzx $ \KwAnd $\Bxx[k]\neq \evenx$}\label{line:B=0hxxB2even}
      \State $\bid\gets k$\Comment{A new block of $\bv{h-1}$ is starting}\label{line:blockstartB2}
    \EndIf
    \If{$\Bxx[k] = \oddx$}\label{line:B=0hxxB2odd}
      \State $\Bxx\gets \oox$ \Comment{{Mark the block as old}}
    \EndIf
\item[]$\vdots$
\setcounter{ALG@line}{12}
      \If{$\Bxx[j] = \zzx$} \Comment{Check if already marked}\label{line:new_startB2}
        \State$\Bxx[j] \gets \evenx$\Comment{A new block of $\bv{h}$ will start here}\label{line:writehB2}
      \EndIf
\end{algorithmic}
\smallbreak\hrule
\caption{Modification of the \hm\ algorithm to use a two-bit array $\Bxx$ instead of the integer array $B$. The code shows the case for $h$ even; if $h$ is odd, the value $\evenx$ is replaced by $\oddx$ and viceversa.}\label{fig:HMBxx}
\end{figure}

Notice that, by Corollary~\ref{cor:lcp}, the value in $\Bxx[i]$ changes from $\zzx$ to $\evenx$ or $\oddx$ during iteration $h=\lcp_{01}[i]+1$. Hence, if every time we do such change we write to an external file the pair $\langle i, h-1 \rangle$, when the merging is complete the file contains all the information required to compute the LCP array $\lcpzo$ even if we do not know $\lcpz$ and $\lcpo$. This idea has been introduced and investigated in~\cite{wabi/EgidiLMT18}.

\textbf{}

\newcommand{\Tz}{{T_0}}
\newcommand{\Tone}{{T_1}}
\newcommand{\Tzo}{{T_{01}}}

\newcommand{\ellz}{\mathsf{L}_0}
\newcommand{\Piz}{\Pi_0}
\newcommand{\ello}{\mathsf{L}_1}
\newcommand{\ellj}[1]{\mathsf{L}_{#1}}
\newcommand{\Pio}{\Pi_1}

\newcommand{\hei}{\mathsf{hgt}}
\newcommand{\avehei}{\mathsf{avehgt}}
\newcommand{\Cxx}{C_2}

\section{Merging compressed tries}\label{sec:tries}

\renewcommand{\xx}{\mbox{\texttt{\#}}}

\begin{figure}[t]
\includegraphics[width=\textwidth]{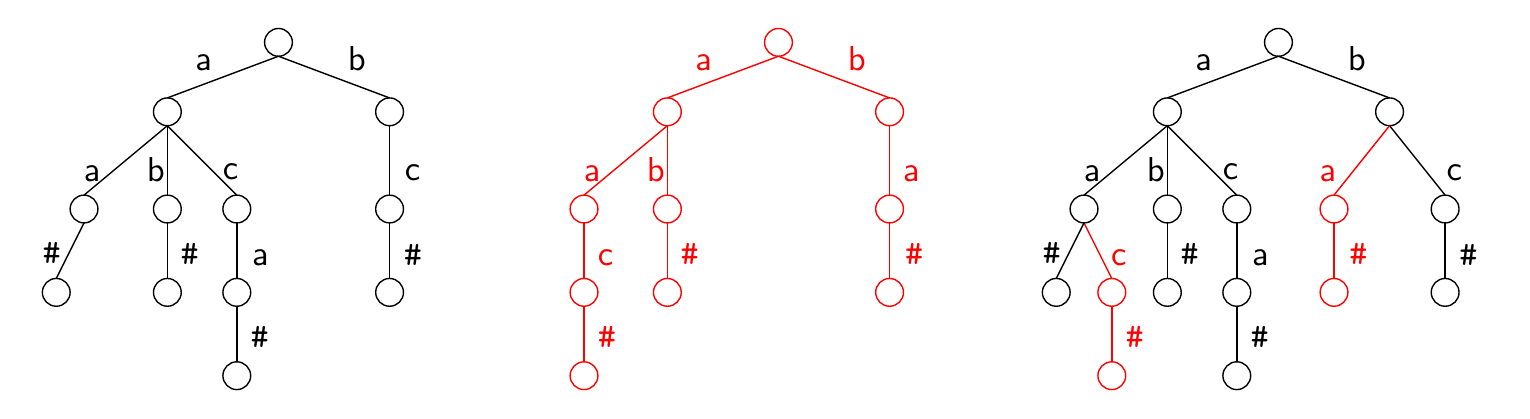}

\begin{center}
\sf\footnotesize
\begin{tabular}[t]{c@{\hskip 10mm}c@{\hskip 12mm}c}
\begin{tabular}[t]{|c|c|l|}\hline
$\last_0$& ~$\ellz$~ & $\Piz$\\ \hline
  0 &  a & \eps  \\
  1 &  b &  \\

  0 &  a & a \\
  0 &  b &  \\
  1 &  c &  \\

  1 &\xx & aa \\

  1 &\xx & aca  \\

  1 &  c &  b   \\

  1 &\xx & ba  \\
  1 &  a & ca  \\
  1 &\xx & cb\\  \hline
\end{tabular}&
\begin{tabular}[t]{|c|c|l|}\hline
$\last_1$& ~$\ello$~ & $\Pio$\\ \hline
  0 &  a & \eps  \\
  1 &  b &  \\

  0 &  a & a \\
  1 &  b &  \\

  1 &  c & aa \\

  1 &\xx & ab   \\

  1 &  a & b  \\

  1 &\xx & ba  \\
  1 &\xx & caa \\
  \hline
\end{tabular}
&
\begin{tabular}[t]{|c|c|l|}\hline
$\last_{01}$& ~$\ellj{01}$~ & $\Pi_{01}$\\ \hline
  0 &  a & \eps  \\
  1 &  b &  \\

  0 &  a & a \\
  0 &  b &  \\
  1 &  c &  \\

  0 &\xx & aa \\
  1 &  c &  \\

  1 &\xx & ab   \\
  1 &\xx & aca  \\

  0 &  a & b  \\
  1 &  c &    \\

  1 &\xx & ba  \\
  1 &  a & ca  \\
  1 &\xx & caa \\
  1 &\xx & cb\\  \hline
\end{tabular}
\end{tabular}
\end{center}

\caption{The trie $\Tz$ containing the strings {\sf aa\xx, ab\xx, aca\xx, bc\xx} (left), the trie $\Tone$ containing {\sf aac\xx, ab\xx, ba\xx} (center) and the trie $\Tzo$ containing the union of the two set of strings (right). Below each trie we show the corresponding XBWT representation.\label{fig:tries}}
\end{figure}

Tries~\cite{knuth3} are a fundamental data structure for representing a collection of $k$ distinct strings. A trie consists of a rooted tree in which each edge is labeled with a symbol in the input alphabet, and each string is represented by a path from the root to one of the leaves. To simplify the algorithms, and ensure that no string is the prefix of another one, it is customary to add a special symbol $\xx \not\in \A$ at the end of each string.\footnote{In this and in the following section we purposely use a special symbol~\xx\ different from~\$. The reason is that \$ is commonly used to for sorting purposes, while~\xx\ simply represents a symbol different from the ones in~$\A$.} Tries for different sets of strings are shown in Figure~\ref{fig:tries}. For any trie node $u$ we write $\hei(u)$ to denote its height, that is the length of the path from the root to $u$. We define the height of the trie $T$ as the maximum node height $\hei(T)=\max_u \hei(u)$, and the average height $\avehei(T) = (\sum_u \hei(u))/|T|$, where $|T|$ denotes the number of trie nodes.

The eXtended Burrows-Wheeler Transform~\cite{jacm09,spire/Manzini16,spire/OhlebuschSB18} is a generalization of the BWT designed to compactly represent any labeled tree~$T$. To define $\XBWT(T)$,  to each {\em internal} node $w$ we associate the string $\lambda_w$ obtained by concatenating the symbols in the edges in the upward path from $w$ to the root of $T$. If $T$ has $n$ internal nodes we have $n$ strings overall; let $\Pi[1,n]$ denote the array containing such strings sorted lexicographically. Note that $\Pi[1]$ is always the empty string corresponding to the root of $T$.
For $i=1,\ldots,n$ let $L(i)$ denote the set of symbols labeling the edges exiting from the node corresponding to $\Pi[i]$. We define the array $\ellj{}$ as the concatenation of the arrays $L(1), \ldots, L(n)$. If $T$ has $m$ edges (and therefore $m+1$ nodes), it is $|\ellj{}|=m$ and  $\ellj{}$ contains $n-1$ symbols from $\A$ and $m+1-n$ occurrences of~\xx. To keep an explicit representation of the intervals $L(1), \ldots, L(n)$ within $\ellj{}$, we define a binary array $\last[1,m]$ such that $\last[i]=\onex$ iff $\ellj{}[i]$ is the last symbol of some interval $L(j)$. See Figure~\ref{fig:tries} for a complete example.

In~\cite{xbw05} it is shown that the two arrays $\XBWT(T) = \langle \last, \ellj{} \rangle$ are sufficient to represent~$T$, and that if they are enriched with data structures supporting constant time rank and select operations, $\XBWT(T)$ can be used for efficient upward and downward navigation and for substring search in~$T$. The fundamental property for efficient navigation and search is that there is an one-to-one correspondence between the symbols in $\ellj{}$ different from $\xx$ and the strings in $\Pi$ different from the empty string. The correspondence is order preserving in the sense that the $i$-th occurrence of symbol $c$ corresponds to the $i$-th string in $\Pi$ starting with $c$. For example, in Figure~\ref{fig:tries} (right) the third {\sf a} in $\last_{01}$ corresponds to the third string in $\Pi_{01}$ starting with {\sf a}, namely {\sf ab}. Note that {\sf ab} is the string associated to the node reached by following the edge associated to the third {\sf a} in $\ellj{01}$.

In this section, we consider the problem of merging two distinct XBWTs. More formally, let $\Tz$ (resp. $\Tone$) denote the trie containing the set of strings $\tj{1},\ldots, \tj{k}$ (resp. $\sj{1},\ldots, \sj{h}$), and let $\Tzo$ denote the trie containing the strings in the union $\tj{1}$,\ldots, $\tj{k}$, $\sj{1}$, \ldots, $\sj{h}$ (see Figure~\ref{fig:tries}). Note that $\Tzo$ might contain less than $h+k$ strings: if the same string appears in both $\Tz$ and $\Tone$ it will be represented in $\Tzo$ only once. Given $\xbwt(\Tz) = \langle \lastz, \ellz \rangle$ and $\xbwt(\Tone) = \langle \lasto, \ello \rangle$ we want to compute the XBWT representation of the trie $\Tzo$.

\ignore{In this section we consider the problem of computing the XBWT representation of $\Tzo$ given the XBWT representations $\xbwt(\Tz) = \langle \lastz, \ellz \rangle$ and $\xbwt(\Tone) = \langle \lasto, \ello \rangle$.}

We observe that if we had at our disposal the sorted string arrays $\Piz$ and $\Pio$, then the construction of $\xbwt(\Tzo)$ could be done as follows: First, we merge lexicographically the strings in $\Piz$ and $\Pio$, then we scan the resulting sorted array of strings. During the scan 
\begin{itemize}
    \item if we find a string appearing only once then it corresponds to an internal node belonging to either $\Tz$ or $\Tone$; the labels on the outgoing edges can be simply copied from the appropriate range of $\ellz$ or $\ello$.

    \item if we find two consecutive equal strings they correspond respectively to an internal node in $\Tz$ and to one in $\Tone$. The corresponding node in $\Tzo$ has a set of outgoing edges equal to the union of the edges of those nodes in $\Tz$ and $\Tone$: thus, the labels in the outgoing edges are the union of the symbols in the appropriate ranges of $\ellz$ and $\ello$. 
\end{itemize}

Although the arrays $\Piz$ and $\Pio$ are not available, by properly modifying the \hm\ algorithm we can compute how their elements would be interleaved by the merge operation. Let $\mz = |\ellz|=|\lastz|$, $\nz = |\Piz|$, and similarly $\mone = |\Lxo|=|\lasto|$, $\none = |\Pio|$. 
Fig.~\ref{fig:xHMalgo} shows the code for the generic $h$-th iteration of the \hm\ algorithm adapted for the XBWT. Iteration $h$ computes a binary vector $\bv{h}$ containing $n_0=|\tz|$ \zerob's and $n_1 = |\tone|$ \oneb's and such that the following property holds (compare with Property~\ref{prop:hblock})

\begin{figure}
\hrule\smallbreak
\begin{algorithmic}[1]
\State Initialize array $F[1,\sigma]$
\State $k_0 \gets 1$; $k_1 \gets 1$  \Comment{Init counters for $\ellz$ and $\ello$}
\State $\bv{h} \gets \zerob\oneb$  \Comment{First two entries correspond to $\Piz[1]=\Pio[1]=\epsilon$}\label{step:01}
\For{$k \gets 1$ \KwTo $n_0+n_1$}
    \State $b \gets \bv{h-1}[k]$\Comment{Read bit $b$ from $\bv{h-1}$}
    \Repeat{}
      \State $c \gets \ellj{b}[k_b]$  
      \Comment{Get symbol from $\ellz$ or $\ello$ according to $b$}
      \If {$c \neq \xx$} \Comment{$\xx$ is ignored: it is not in $\Piz$ or $\Pio$}
        \State $j \gets F[c]{\mathsf ++}$ \Comment{Get destination for $b$ according to symbol $c$}
        \State $\bv{h}[j] \gets b$        \Comment{Copy bit $b$ to $\bv{h}$}\label{line:updatebv}
      \EndIf
      \State $\ell \gets \lastj{b}[k_b{\mathsf ++}]$ \Comment{Check if $c$ labels last outgoing edge}
    \Until{$\ell\neq 1$}
\EndFor
\end{algorithmic}
\smallbreak\hrule
\caption{Main loop of algorithm \hm\ modified to merge XBWTs. Array $F$ is initialized so that $F[c]$ contains the number of occurrences of symbols smaller than $c$ in $\ellz$ and $\ello$ plus three, to account for $\Piz[1]=\Pio[1]=\epsilon$ which are smaller than any other string.}\label{fig:xHMalgo}
\end{figure}

\begin{property}\label{prop:xhblock}
At the end of iteration~$h$, for $i=2,\ldots, \nz$ and $j=2,\ldots \none$ the $i$-th \zerob\ precedes the
$j$-th \oneb\ in $\bv{h}$ if and only if 
\begin{equation}\label{eq:xhblock}
\Piz[i][1,h] \;\preceq\; \Pio[j][1,h].
\end{equation}\qed
\end{property}

In~\eqref{eq:xhblock} $\Piz[i][1,h]$ denotes the length-$h$ prefix of $\Piz[i]$. If $\Piz[i]$ has length smaller than $h$ then $\Piz[i][1,h]=\Piz[i]$ (and similarly for $\Pio[j]$). 
Note that Property~\ref{prop:xhblock} does not mention the first $\zerob$ and the first $\oneb$ in $\bv{h}$: By construction it is $\Piz[1]=\Pio[1]=\epsilon$ so we know their lexicographic rank is the smallest possible. Note also that because of Step~\ref{step:01} in Fig.~\ref{fig:xHMalgo}, the first $\zerob$ and the first $\oneb$ in~$\bv{h}$ are always the first two elements of $\bv{h}$. 

Apart from the first two entries, during iteration $h$ the array $\bv{h}$ is logically partitioned into $\sigma$ subarrays, one for each alphabet symbol different from~$\xx$. If $Occ(c)$ denotes the number of occurrences in $\ellz$ and $\ello$ of the symbols smaller than $c$, then the subarray corresponding to $c$ starts at position $Occ(c)+3$. Hence, if $c<c'$ the subarray corresponding to $c$ precedes the one corresponding to $c'$. Because of how the array $F$ is initialized and updated, we see that every time we read a symbol $c$ from $\ellz$ and $\ello$ we write a value in the portion of $\bv{h}$ corresponding to $c$, and that each portion is filled sequentially. Armed with these observations, we are ready to establish the correctness of the algorithm in Figure.~\ref{fig:xHMalgo}. 

\begin{lemma}\label{lemma:xhblock}
Let $\bv{0} = \zerob\oneb\zerob^{n_0-1}\oneb^{n_1-1}$, and let $\bv{h}$ be obtained from $\bv{h-1}$ by the algorithm in Fig.~\ref{fig:xHMalgo}. Then, for $h=0,1,2,\ldots$, the array $\bv{h}$ satisfies Property~\ref{prop:xhblock}.
\end{lemma}

\begin{proof}
We prove the result by induction. For $h=0$,
$\Piz[i][1,0]=\Pio[j][1,0]=\epsilon$ so~\eqref{eq:xhblock} is always true and $\bv{0}$ satisfies Property~\ref{prop:xhblock}.

Suppose now $h>0$. To prove the ``if'' part, let $3 \leq v < w \leq \nz+\none$ denote two indexes such that $\bv{h}[v]$ is the $i$-th \zerob\ and $\bv{h}[w]$ is the $j$-th \oneb\ in $\bv{h}$ for some $2 \leq i \leq \nz$ and $2 \leq j \leq \none$ (it is $v \geq 3$ since $i\geq2$ and $\bv{h}[1,2]=\zerob\oneb$). We need to show that~\eqref{eq:xhblock} holds.

Assume first $\Piz[i][1]\neq \Pio[j][1]$. The hypothesis $v<w$ implies $\Piz[i][1]< \Pio[j][1]$ hence~\eqref{eq:hblock} certainly holds. Assume now $\Piz[i][1] =  \Pio[j][1] = c$. Let $v'$, $w'$ denote respectively the values of the main loop variable~$k$ in the procedure of Figure~\ref{fig:xHMalgo} when the entries $\bv{h}[v]$ and $\bv{h}[w]$ are written (hence, during the scanning of $\bv{h-1}$). The hypotheses $v<w$ and $\Piz[i][1]= \Pio[j][1]$ imply ${v}' < {w}'$. By construction
$\bv{h-1}[{v}']=\zerox$ and $\bv{h-1}[{w}']=\onex$. Say ${v}'$ is the $i'$-th \zerob\ in $\bv{h-1}$ and ${w}'$ is the $j'$-th \oneb\ in $\bv{h-1}$. By the inductive hypothesis on $\bv{h-1}$ we have
\begin{equation}\label{eq:xhblock2}
\Piz[i'][1,h-1] \;\preceq\; \Pio[j'][1,h-1]
\end{equation}
(we could have $v'=1$ that would imply $i'=1$; in that case we cannot apply the inductive hypothesis, but~\eqref{eq:xhblock2} still holds). By the properties of the XBWT we have 
$$
\Piz[i][1,h] = c\,\Piz[i'][1,h-1]\qquad\mbox{and}\qquad
\Pio[j][1,h] = c\,\Pio[j'][1,h-1]
$$
which combined with~\eqref{eq:xhblock2} gives us~\eqref{eq:xhblock}. 

For the ``only if'' part assume~\eqref{eq:xhblock} holds for some $i\geq 2$ and $j\geq 2$. We need to prove that in $\bv{h}$ the $i$-th \zerob\ precedes the $j$-th \oneb. If $\Piz[i][1]\neq \Pio[j][1]$ the proof is immediate. If $c=\Piz[i][1]=\Pio[j][1]$ then
$$
\Piz[i][2,h]\preceq \Pio[j][2,h].
$$
Let $i'$ and $j'$ be such that $\Piz[i'][1,h-1] = \Piz[i][2,h]$ and $\Pio[j'][1,h-1] = \Pio[j][2,h]$.
By induction, in $\bv{h-1}$ the $i'$-th \zerob\ precedes the $j'$-th \oneb\ (again we could have $i'=1$ and in that case we cannot apply the inductive hypothesis, but the claim still holds). 

During iteration~$h$, the $i$-th \zerob\ in $\bv{h}$ is written to position $v$ when processing the $i'$-th \zerob\ of $\bv{h-1}$, and the $j$-th \oneb\ in $\bv{h}$ is written to position $w$ when processing the $j'$-th \oneb\ of $\bv{h-1}$. Since in $\bv{h-1}$ the $i'$-th \zerob\ precedes the $j'$-th \oneb\ and since $v$ and $w$ both belongs to the subarray of $\bv{h}$ corresponding to the symbol $c$, their relative order does not change and the $i$-th \zerob\ precedes the $j$-th \oneb\ as claimed.\qed
\end{proof}

As in the original \hm\ algorithm we stop the merge phase after the first iteration $h$ such that $\bv{h} = \bv{h-1}$. Since in subsequent iterations we would have $\bv{g} = \bv{h}$ for any $g>h$, we get that by Property~\ref{prop:xhblock}, $\bv{h}$ gives the correct lexicographic merge of $\Piz$ and $\Pio$. Note however that the lexicographic order is not sufficient to establish whether two consecutive nodes, say $\Piz[i]$ and $\Pio[j]$ have the same upward path and therefore should be merged in a single node of~$\Tzo$. To this end, we consider the integer array~$B$ used in Section~\ref{sec:hem} to mark the starting point of each block. We have shown in Corollary~\ref{cor:lcp} that at the end of the original \hm\ algorithm $B$ contains the LCP values plus one. Indeed, at iteration $h$ the algorithm sets $B[k]=h$ since it ``discovers'' that the suffixes in $\sazo[k-1]$ and $\sazo[k]$ differ in the $h$-th symbol (hence $\lcpzo[k]=h-1$). If we maintain the array $B$ in the XBWT merging algorithm, we get that at the end of the computation if the strings associated to  $\Piz[i]$ and $\Pio[j]$ are identical then the entry in $B$ corresponding to $\Pio[j]$ would be zero, since the two strings do not differ in any position. Hence, at the end of the modified \hm\ algorithm the array $\bv{h}$ provides the lexicographic order of the nodes, and the array $B$ the position of the nodes of $\Tz$ and $\Tone$ with the same upward path.  We conclude that with a single scan of $\bv{h}$ and $B$ we can merge all paths and compute $\xbwt(\Tz)$. Finally, we observe that instead of $B$ we can use a two-bit array $\Bxx$ as in Sect.~\ref{sec:B2}, since we are only interested in determining whether a certain entry is zero, and not in its exact value.

\begin{lemma} \label{lemma:hmx}
The modified \hm\ algorithm computes $\xbwt(\Tzo)$ given $\xbwt(\Tz)$ and $\xbwt(\Tone)$ in $\Oh( |\Tzo|\hei(\Tzo))$ time and $4n + \Oh(\log n)$ bits of working space, where $n=\nz+\none$. 
\end{lemma}

\begin{proof}
Each iteration of the merging algorithm takes $\Oh(\mz+\mone)$ time since it consists of a scan of the arrays $\bv{h-1}$, $\ellz$, $\ello$, $\lastz$ and $\lasto$. After at most $\hei(\Tzo)$ iterations the strings in $\Piz$ and $\Pio$ are lexicographically sorted and $\bv{h}$ no longer changes. The final scan of $\bv{h}$ and $\Bxx$ to compute $\xbwt(\Tz)$ takes $\Oh(\mz+\mone)$ time. Since $|\Tzo| \geq \max(\mz,\mone)$ the overall cost is $\Oh(|\Tzo|\hei(\Tzo))$ time.
The working space of the algorithm, consists of $\Bxx$ and of two instances of the $\bv{h}$ array (for the current and the previous iteration), in addition to $\Oh(\sigma)$ counters (recall that $\sigma$ is assumed to be constant).\qed
\end{proof}

As for BWT/LCP merging, we now show how to reduce the running time by skipping the portions of $\bv{h}$ that no longer change from one iteration to the next. Note that we cannot use monochrome blocks to early terminate XBWT merging. Indeed, from the previous discussion we know that if two strings $\Piz[i]$ and $\Pio[j]$ are equal, they will form a non-monochrome block that will never be split. 

For this reason we introduce an array $C[1,\nz+\none]$ that, at the beginning of iteration $h$, keeps track of all the strings in $\Piz$ and $\Pio$ that have length less than $h$. More precisely, for $i=1,\ldots,\nz$ (resp. $j=1,\ldots,\none$) if the $i$-th \zerob\ (resp. the $j$-th \oneb) is in position $k$ of $\bv{h}$, then $C[k]=\ell>0$ iff the length $|\Piz[i]|$ (resp. $|\Pio[j]|$) is equal to $\ell-1$ with $\ell-1 < h$. As a consequence, if $C[k]=0$ then the string corresponding to $C[k]$ has length $h$ or more. 
Note that by Property~\ref{prop:xhblock} at the beginning of iteration $h$ the algorithm has already determined the lexicographic rank of all the strings in $\Piz$ and $\Pio$ of length smaller than $h$. 
Hence, the entry in $\bv{h}[k]$ will not change in successive iterations and will remain associated to the same string from $\Piz$ or $\Pio$. 

The array $C$ is initialized as $110^{\nz+\none-2}$ since at the beginning of iteration~1 it is $\bv{h}=\zerob\oneb\zerob^{\nz-1}\oneb^{\none-1}$ and indeed the only strings of length 0 are $\Piz[0]=\Pio[0]=\epsilon$. During iteration $h$, we update $C$ adding, immediately after Line~\ref{line:updatebv} in Fig.~\ref{fig:xHMalgo}, the line
$$
\mathbf{if\ } C[k]=h \mathbf{\ then\ } C[j] \gets h+1\qquad
$$
The rationale is that if, during iteration~$h-1$ we found out that the string $\alpha$ corresponding to $\bv{h-1}[k]$ has length $h-1$ (so we set $C[k]=h$), then the string corresponding to $\bv{h}[j]$ is $c\alpha$ and has therefore length~$h$.

By the above discussion we see that if at iteration $h$ we write $h+1$ to position $C[j]$, then at iteration $h+1$ we can possibly use $C[j]$ to write $h+2$ in some other position in $C$, but starting from iteration $h+2$ it is no longer necessary to process neither $C[j]$ nor $\bv{h+2}[j]$ since they will not affect neither $C$ nor $\bv{h+3}$. In other words, during iteration $h$ we can skip all ranges $\bv{h}[\ell,m]$ such that $C[\ell,m]$ contains only positive values smaller than~$h$. These ranges grown larger and larger as the algorithm proceeds and are handled in the same way as the \useless\ blocks in \gap. Finally, we observe that, using the same techniques as in Section~\ref{sec:B2}, we can replace the integer array $C$ with an array $\Cxx$ containing only two bits per entry.

\ignore{therefore reducing the overall working space to $4(\nz+\none)$ bits (the arrays $\bv{h-1}$, $\bv{h}$, and $\Cxx$) plus $\Oh(\sigma)$ words for the array $F$ and other counters, plus the space used for handling the list of \useless\ blocks.}

\begin{theorem}\label{theo:gapx}
The modified \gap\ algorithm computes $\xbwt(\Tzo)$ given $\xbwt(\Tz)$ and $\xbwt(\Tone)$ in $\Oh(|\Tzo| \avehei(\Tzo))$ time. The working space is $6n  + \Oh(\log n)$ bits, where $n=\nz+\none$,  plus the space required for handling irrelevant blocks.  
\end{theorem}

\begin{proof}
The analysis is similar to the one in Theorem~\ref{theo:gap}.
Here the algorithm executes $\hei(\Tzo)$ iterations; however, because of irrelevant blocks, iterations have decreasing costs. To bound the overall running time, observe that the cost of each iteration is dominated by the cost of processing the entries in $\ellz$ and $\ello$. The generic entry $\ellz[i]$ corresponds to a trie node $u_i$ with upward path of length $\hei(u_i)$. Entry $\ellz[i]$ is processed when the \gap\ algorithm reaches the entry in $\bv{h}$ corresponding to the string $\Pio[i']$ associated to $u_i$'s parent. We know that $\bv{h}$'s entry corresponding to $\Pio[i']$ becomes irrelevant after iteration $|\Pio[i']| + 1 = \hei(u_i)$. Hence, the overall cost of processing $u_i$ is $\Oh(\hei(u_i))$. Summing over all entries in $\ellz$ and $\ello$ the total cost is $\Oh(|\Tz|\avehei(\Tz) + |\Tone|\avehei(\Tone))$. The thesis follows observing that $|\Tzo| \avehei(\Tzo) \geq \max\bigl(|\Tz|\avehei(\Tz),|\Tone|\avehei(\Tone)\bigr)$.\qed
\end{proof}


\def\clcpzo{\mathsf{clcp}_{01}}
\def\clcp{\mathsf{cLCP}}

\def\lenghzo{\mathsf{length}_{01}}
\def\lenghz{\mathsf{length}_{0}}
\def\lengho{\mathsf{length}_{1}}

\def\avelen{\mathsf{aveLen}}
\def\maxlen{\mathsf{maxLen}}

\def\aveclcp{\mathsf{avecLcp}}
\def\maxclcp{\mathsf{maxcLcp}}

\def\xxy{\mbox{\texttt{\#}}}
\def\bvpair#1#2{\langle #1, #2\rangle}

\section{Merging indices for circular patterns}\label{zirrs}

Another well known variant of the BWT is the {\em multistring circular BWT} which is defined by sorting the cyclic rotations of the input strings instead of their suffixes. However, to make the transformation reversible, the cyclic rotations have to be sorted according to an order relation, different from the lexicographic order, that we now quickly review.  

For any string $\txx{}$, we define the {\em infinite form} $\txx{}^\infty$ of $\txx{}$ as the infinite length string obtained concatenating $\txx{}$ to itself infinitely many times. Given two strings $\txx{}$ and $\sxx$ we write $\txx{}  \preceq^\infty \sxx$ to denote that $\txx{}^\infty \preceq\sxx^\infty$. For example, for $\txx{}= \mathsf{abaa}$ and $\sxx=\mathsf{aba}$, it is $\txx{}^\infty = \mathsf{abaaabaa\cdots}$ and $\sxx^\infty = \mathsf{abaabaaba}\cdots$ so $\txx{}  \preceq^\infty \sxx$.  Notice that $\txx{}^\infty = \sxx^\infty$ does not necessarily imply that $\txx{} = \sxx{}$. For example, for $\txx{} = \mathsf{ababab}$ and $\sxx{} = \mathsf{abab}$ it is $\txx{}^\infty = \sxx^\infty$. The following lemma, which is a consequence of Fine and Wilf Theorem~\cite{FW65} and a restatement of Proposition~5 in~\cite{tcs/MantaciRRS07}, provides an upper bound to the number of comparisons required to establish whether $\txx{}^\infty = \sxx^\infty$.

\begin{lemma}\label{lemma:limit}
If   $\txx{}^\infty \neq \sxx^\infty$ then there exists an index $i\leq |\txx{}| + |\sxx{}| - \gcd(|\txx{}|,|\sxx{}|)$ such that $\txx{}^\infty[i] \neq \sxx{}^\infty[i]$.\qed
\end{lemma}

A string is {\em primitive} if all its cyclic rotations are distinct. The following Lemma is another well known consequence of the Fine and Wilf Theorem.

\begin{lemma}\label{lemma:primitive}
If $\txx{}$ and $\sxx{}$ are primitive, $\txx{}^\infty = \sxx{}^\infty$ implies $\txx{} = \sxx{}$.\qed
\end{lemma}

Let $\tz[1,\nz],\tone[1,\none]$ be two primitive strings and $\tzo[1,n]$ their concatenation of length $n=\nz + \none$. For $i=1,\ldots, n$, let $\rotzo{i}$ define the rotation of substrings $\tz$ and $\tone$ within $\tzo$ as follows:
$$
\rotzo{i} = \begin{cases}
\tz[i,\nz]\tz[1,i-1] & \mbox{if } 0<i\leq \nz\\
\tone[i-\nz,\none]\tone[1,i-\nz-1] & \mbox{if } \nz<i\leq \nz + \none.
\end{cases}
$$
For example, if $\tz = \mathsf{abc}$ and $\tone = \mathsf{abbb}$, it is $\rotzo{2}= \mathsf{bca}$ and $\rotzo{7}=\mathsf{babb}$. The above definition of rotations of substrings can be obviously generalized to a collection of $k$ strings. 

\ignore{In particular, for $k=1$, it is $\rotz{i} = \tz[i,\nz]\tz[1,i-1]$ and $\roto{i} = \tone[i,\none]\tone[1,i-1]$.}

In addition to assuming that $\tz$ and $\tone$ are primitive, we assume that $\tz$ is not a rotation of $\tone$. We define the {\em circular Suffix Array} of $\tz$ and $\tone$, $\csazo$ as the permutation of $[1,n]$ such that:
\begin{equation}\label{eq:csa}
\rotzo{\csazo[i]} \; \preceq^\infty\; \rotzo{\csazo[i+1]}.
\end{equation}
Note that because of our assumptions and Lemma~\ref{lemma:primitive}, the inequality in~\eqref{eq:csa} is always strict.  Finally, the multistring {\em circular Burrows-Wheeler Transform (cBWT)} is defined as 
\[
\cbwt_{01}[i] =
\begin{cases}
\tz[\nz]                      & \mbox{if } \csazo[i] = 1\\
\tz[\csazo[i] - 1]       & \mbox{if } 1 < \csazo[i] \leq \nz\\
\tone[\none]                    & \mbox{if } \csazo[i] = \nz + 1\\
\tone[\csazo[i]-\nz - 1] & \mbox{if } \csazo[i]> \nz+1.
\end{cases}
\]
The above definition given for $\tz$ and $\tone$ can be generalized to any number of strings. The $\preceq^\infty$ order and the above multistring circular BWT has been introduced in~\cite{tcs/MantaciRRS07}. In~\cite{talg/FerraginaV10} the authors uses a data structure equivalent to a circular BWT to design a {\em compressed permuterm index} for prefix/suffix queries. The crucial observation is that if we add a unique symbol~\xxy\ at the end of each string, the same symbol for every string, then searching $\beta\xxy\alpha$ in a circular BWT returns all the strings prefixed by $\alpha$ and suffixed by~$\beta$. 
In~\cite{Honetal11} Hon et al. use the circular BWT to design a succinct index for circular patterns. Note that Hon et al. in addition to $\cbwtzo$ use an additional data structure $\lenghzo$ such that $\lenghzo(i)$ provides the length of the string $\txx{j}$ to which the symbol $\cbwtzo[i]$ belongs. 
Finally, a lightweight algorithm for the construction of the circular BWT has been described in~\cite{cpm/HonKLST12}: for a string of length~$n$ the proposed algorithm takes $\Oh(n)$ time and uses $\Oh(n\log\sigma)$ bits of space.

To simplify our analysis, we preliminary extend the concept of longest common prefix to the $\preceq^\infty$ order. For any pair of strings $\txx{}, \sxx{}$ we define
\begin{equation}\label{eq:cLCP}
\clcp(\txx{}, \sxx{}) = 
\begin{cases}
\LCP(\txx{}^\infty, \sxx{}^\infty) & \mbox{if } \txx{}^\infty \neq \sxx{}^\infty\\
|\txx{}| +  |\sxx{}| - \gcd(|\txx{}|,|\sxx{}|)  & \mbox{otherwise}.
\end{cases}
\end{equation}
Because of Lemma~\ref{lemma:limit}, $\clcp(\txx{}, \sxx{})$ generalizes the standard LCP in that it provides the number of comparisons that are necessary in order to establish the $\preceq^\infty$ ordering between $\txx{}$, $\sxx{}$. It is then natural to define
for $i=2,\ldots,n$
\begin{equation}\label{eq:clcp}
\clcpzo[i] = \clcp(\rotzo{\csazo[i-1]},\rotzo{\csazo[i]})
\end{equation}
and the values
\begin{equation}
\maxclcp = \max\nolimits_i \clcpzo[i], \qquad 
\aveclcp = \bigl(\sum\nolimits_i \clcpzo[i]\bigr)/n.
\end{equation}
that generalize the standard notions of maximum LCP and average LCP.

\ignore{Lo  vogliamo mettere? Note it is $\maxclcp \leq 2\; \maxlen$ and $\aveclcp \leq 2\; \bigl( \sum_i |\rotzo{i}|\bigr)/n$.}

Let $\cbwtz$ (resp. $\cbwto$) denote the circular BWT for the collection of strings 
$\tj{1},\ldots, \tj{k}$ (resp. $\sj{1},\ldots, \sj{h}$). In this section we consider the problem of computing the circular BWT $\cbwtzo$ for the union collection $\tj{1}$,\ldots, $\tj{k}$, $\sj{1}$, \ldots, $\sj{h}$. As we previously observed, we assume that all strings are primitive and that within each input collection no string is the rotation of another. However, we cannot rule out the possibility that some $\tj{i}$ is the rotation of some $\sj{j}$. The merging algorithm should therefore recognize this occurrence and eliminate from the union one of the two strings, say $\sj{j}$. In practice, this means that all symbols of $\cbwto$ coming from $\sj{j}$ must not be included in~$\cbwtzo$. 

To merge $\cbwtz$ and $\cbwto$ we need to merge their symbols according to their context. By construction, the context of $\cbwtz[i]$ (resp. $\cbwto[j]$) is $\rotz{\csaz[i]}$ (resp. $\roto{\csao[j]}$), where $\rotz{\csaz[i]}$ is a cyclic rotation of 
the string $\tj{k}$ to which the symbol $\cbwtz[i]$ belongs (and similarly for $\roto{\csao[j]}$). Note however, that context must be sorted according to the $\prec^\infty$ order; hence $\cbwtz[i]$ should precede $\cbwto[j]$ in $\cbwtzo$ iff  $\rotz{\csaz[i]} \preceq^\infty \roto{\csao[j]}$. The good news is that the \hm\ algorithm, as described in Figure~\ref{fig:HMalgo}, when applied to $\cbwtz$ and $\cbwto$ will sort each symbol according to the $\preceq^\infty$ order of its context. Notice that the $\preceq^\infty$ order induces a significant difference with respect to the merging of  BWTs: indeed, since there are no \eos's in $\cbwtz$ and $\cbwto$ Line~\ref{line:moveeos} is never executed and the destination of each symbol is always determined by its predecessor in the cyclic rotation. More formally, reasoning as in Lemma~\ref{lemma:hblock}, it is possible to prove the following property. 

\begin{property}\label{prop:chblock}
For $i=1,\ldots, n_0$ and $j=1,\ldots n_1$ the $i$-th \zerob\ precedes the
$j$-th \oneb\ in $\bv{h}$ if and only if
\begin{equation}\label{eq:chblock}
\rotz{\csa_0[i]}^\infty[1, h] \; \preceq\; \roto{\csa_1[j]}^\infty[1,h].
\end{equation}\qed
\end{property}

Property~\ref{prop:chblock} states that after iteration~$h$ the infinite strings $\rotz{\csa_0[i]}^\infty$ and $\roto{\csa_1[j]}$ have been sorted according to their length $h$ prefix. As for the original \hm\ algorithm, as soon as $\bv{h+1}=\bv{h}$ the $\bv{\cdot}$ array will not change in any successive iteration and the merging is complete. 
By Lemma~\ref{lemma:limit} it is $\bv{h+1}=\bv{h}$ for some $h\leq \maxclcp$.

Since we do not simply need to sort the context, but also 
recognize if some string $\tj{i}$ is a rotation of some $\sj{j}$, we make use of the algorithm in Figure~\ref{fig:HMlcp} which, in addition to $\bv{h}$, also computes the integer array $B$ that marks the boundaries of the groups of all rotations whose infinite form have a common prefix of length~$h$. We can prove a result analogous to Lemma~\ref{lemma:B} replacing the LCP between suffixes ($\lcpzo$) with the LCP between the infinite strings $\rotinf{b}{\csab[i]}$ (that is $\clcpzo$). After iteration $h=\maxclcp$ all distinct rotations have been sorted according to the $\preceq^\infty$ order; thus an entry $B[k]=0$ denotes two rotations $\rotz{\csa_0[i]}^\infty$ and $\roto{\csa_1[j]}^\infty$ which have a common prefix of length $\maxclcp$. By Lemma~\ref{lemma:limit} it is $\rotz{\csa_0[i]}^\infty = \roto{\csa_1[j]}^\infty$ and by Lemma~\ref{lemma:primitive} $\rotz{\csa_0[i]} = \roto{\csa_1[j]}$. The two rotations are therefore identical and the symbol $\cbwto[j]$ should not be included in~$\cbwtzo$. 

Summing up, to merge $\cbwtz$ and $\cbwto$ we execute the procedure of Figure~\ref{fig:HMlcp} until both $\bv{h}$ and $B$ do not change. Then, we compute $\cbwtzo$ by merging $\cbwtz$ and $\cbwto$ according to $\bv{h}$, discarding those symbols corresponding to zero entries in~$B$. The number of iterations will be at most $\maxclcp$. In addition, since we are only interested in zero/nonzero entries, instead of $B$ we can use a 2-bit array $\Bxx$ as in Section~\ref{sec:B2}. Reasoning as for Lemma~\ref{lemma:hmlcp}, setting $n = \nz+\none$ we get the following result.

\begin{lemma} \label{lemma:hmc}
The modified \hm\ algorithm computes $\cbwtzo$ given $\cbwtz$ and $\cbwto$ in $\Oh(  n\,\maxclcp)$ time and $4n + \Oh(\log n)$ bits of working space.\qed
\end{lemma}

As we have done in the previous sections, we now show how to reduce the running time of the merging algorithm by avoiding to re-process the blocks of $\bv{h-1}$ that have become irrelevant for the computation of the new bitarray $\bv{h}$. Reasoning as in Section~\ref{sec:gap} we observe that monochrome blocks, i.e. blocks containing entries only from $\cbwtz$ or $\cbwto$, after having been processed once, become irrelevant and can be skipped in successive iterations. Note however, that whenever $\rotz{\csa_0[i]}^\infty = \roto{\csa_1[j]}^\infty$ these two entries will always belong to the same block. 
To handle this case we first assume $\cbwtzo$ is to be used as a compressed index for circular patterns~\cite{Honetal11} and we later consider the case in which $\cbwtzo$ is to be used for a compressed permuterm index.

\subsection{Compressed indices of circular patterns}

In this setting, $\cbwtzo$ is to be used as a compressed index for circular patterns and therefore we have access to the $\lenghz$ and $\lengho$ data structures providing the length of each rotation.
Under this assumption we modify the \gap\ algorithm described in Section~\ref{sec:gap} as follows: in addition to skipping monotone blocks, every time there is a size-2 non monochrome block containing, say $\cbwtz[i]$ and $\cbwto[j]$, we mark it as {\em quasi-irrelevant} and compute $\ell_{ij} = |\lenghz(i)| + |\lengho(j)| - \gcd(|\lenghz(i)|,|\lengho(j)|)$. As soon as this block is split or we reach iteration $\ell_{ij}$ the block becomes irrelevant and is skipped in successive iterations. As in the original \gap\ algorithm, the computation stops when all blocks have become irrelevant.

For simplicity, in the next theorem we assume that the access to the data structures $\lenghz$ and $\lengho$ takes constant time. If not, and random access to the individual lengths takes $\Oh(\rho)$ time, the overall cost of the algorithm is increased by $\Oh((\nz+\none)\rho)$ since each length is computed at most once. 

\begin{theorem}\label{theo:gapc}
The modified \gap\ algorithm computes $\langle\cbwtzo,\lenghzo\rangle$ given $\langle\cbwtz,\lenghz\rangle$ and $\langle\cbwto,\lengho\rangle$ in $\Oh(n\; \aveclcp)$ time, where $n=\nz+\none$. The working space is $2n + \Oh(\log n)$ bits plus the space required for handling (quasi-)irrelevant blocks.  
\end{theorem}

\begin{proof}
If $\rotzo{\csazo[k]}$ is different from any other rotation, 
by definitions~\eqref{eq:cLCP} and~\eqref{eq:clcp} after at most $\max(\clcpzo[k],\clcpzo[k+1])$ iterations it will be in a 
monochrome (possibly singleton) block. If instead $\rotzo{\csazo[k]}$ is identical to another rotation, which can only be either $\rotzo{\csazo[k-1]}$ or $\rotzo{\csazo[k+1]}$, then after at most 
\begin{equation}\label{eq:max4}
\max(\clcpzo[k-1],\clcpzo[k],\clcpzo[k+1],\clcpzo[k+2])
\end{equation}
iterations it will be in a size-2 non-monochrome block together with its  identical rotation. 
In either case, the block containing $\rotzo{\csazo[k]}$ will become irrelevant and it will be no longer processed in successive iterations. Hence, the overall cost of handling $\rotzo{\csazo[k]}$ over all iterations is proportional to~\eqref{eq:max4}, and the overall cost of handling all rotations is bounded by $\Oh(n\;\aveclcp)$ as claimed.
Note that the final bitarray $\bv{h}$ describes also how $\lenghz$ and $\lengho$ must be interleaved to get $\lenghzo$.\qed 
\end{proof}

\subsection{Compressed permuterm indices}
Finally, we consider the case in which $\cbwtzo$ is to be used as the core of a compressed permuterm index~\cite{talg/FerraginaV10}. In this case we do not have the $\lenghz$ and $\lengho$ data structures, but each string in the collection is terminated by a unique~\xxy\ symbol. In this case, to recognize whether a size-2 non-monochrome block contains two identical rotations, we make use of the following lemma. 

\begin{lemma}\label{lemma:2sharp}
Let $\tx$ and $\sxx{}$ denote two strings each one containing a single occurrence of the symbol~\xxy. If for some $h>0$ it is\;
$
{\tx}^\infty[1,h] = {\sxx{}}^\infty[1,h]
$\;
and\/ ${\tx}^\infty[1,h]$ contains two occurrences of~\xxy, then\/ $\tx=\sxx{}$.
\end{lemma}

\begin{proof}
Let $\delta$ denote the distance between the two occurrences of~\xxy\ in ${\tx}^\infty[1,h]$. Since $\tx$ contains a single~\xxy, we have
$
\tx = {\tx}^\infty[1,\delta] = {\sxx{}}^\infty[1,\delta] = \sxx{}
$.\qed
\end{proof}

The above lemma suggests to design a \cegap algorithm to merge compressed permuterm indices in which the arrays $\bv{\cdot}$ are arrays of pairs so that they keep track also of the number of \xxy~in each prefix. In the following $\bv{h}[k]=\bvpair{b}{m}$ means that the $k$-th rotation belongs to $\csab$, and among the first~$h$ symbols of the infinite form of that rotation there are exactly $m$ occurrences of~\xxy. Formally, for $h=0,1,2,\ldots$ the array $\bv{h}$ satisfies the following property. 

\begin{property}\label{prop:newchblock}
At the end of iteration~$h$ of\/ \cegap Property~\ref{prop:chblock} holds and if $\bv{h}[k]=\bvpair{b}{m}$ is the $i$-th $b$ in $\bv{h}$ then 
$\rotb{\csa_b[i]}^\infty[1,h]$ contains exactly $m$ copies of symbol~\xxy.\qed
\end{property}

Initially we set $\bv{0} = \bvpair{\zerox}{0}^\nz \bvpair{\onex}{0}^\none$ which clearly satisfies Property~\ref{prop:newchblock}. At each iteration  \cegap reads $\bv{h-1}$ and updates $\bv{h}$ using Lines~\ref{line:cegap:block_process_start}--\ref{line:cegap:updatebv} below instead of Lines~\ref{line:block_process_start}--\ref{line:hem:updatebv} of Figure~\ref{fig:HMlcp}:
\begin{center}
\begin{minipage}{11cm}
\begin{algorithmic}[1]\setcounter{ALG@line}{6}
  \State $\bvpair{b}{m} \gets \bv{h-1}[k]$\label{line:cegap:block_process_start}
\State $c \gets \bwt_b[k_b{\mathsf ++}]$ \Comment{Get $c$ according to $b$}
\item[]$\vdots$
\setcounter{ALG@line}{13}
\State $\mathbf{if\ } c=\xxy \mathbf{\ then\ } m \gets m+1$\Comment{Update number of \xxy}\label{line:cegap:updatecount}
\State $\bv{h}[j] \gets \bvpair{b}{m}$\label{line:cegap:updatebv}
\end{algorithmic}
\end{minipage}
\end{center}
Reasoning as in the previous sections, one can prove by induction that with this modification the array~$\bv{h}$ computed by \cegap satisfies Property~\ref{prop:newchblock}. In the \cegap algorithm a block becomes irrelevant when it is monochrome or it is a size-2 non-monochrome block $\bv{h}[k,k+1]$ such that $\bv{h}[k] = \bvpair{\zerox}{2}$ and $\bv{h}[k+1] = \bvpair{\onex}{2}$. By Lemma~\ref{lemma:2sharp} such block corresponds to two identical rotations $\rotz{\csa_0[i]} = \roto{\csa_1[j]}$ and after being processed a final time it can be ignored in successive iterations. 

In the practical implementation of the~\cegap algorithm, instead of maintaining the pairs $\bvpair{b}{m}$, we maintain two bit arrays $\bv{h-1}$, $\bv{h}$ as in \gap, and an additional 2-bit array $C$ containing the second component of the pairs. For such array $C$ two bits per entry are sufficient since the values stored in each entry $C[k]$ never decrease and they are no longer updated when they reach the value~2. 

\begin{theorem}\label{theo:gapc2}
The \cegap algorithm merges two compressed permuterm indices 
$\cbwtz$ and $\cbwto$ in $\Oh(n\; \aveclcp)$ time, where $n=\nz+\none$. The working space is $6n + \Oh(\log n)$ bits plus the space required for handling irrelevant blocks.
\end{theorem}

\begin{proof}
We reason as in the proof of Theorem~\ref{theo:gapc} except that if $\rotzo{\csazo[k]}=\rotzo{\csazo[k+1]}$ we are guaranteed that the corresponding size-2 block will become irrelevant only after iteration $h = 2\, |\rotzo{\csazo[k]}| = 2\, \clcpzo[k+1]$. Hence, the cost of handling $\rotzo{\csazo[k]}$ is still proportional to~\eqref{eq:max4} and the overall cost of the algorithm is  $\Oh(n\; \aveclcp)$ time. The space usage is the same as in Theorem~\ref{theo:gapc}, except for the $2n$ additional bits for the $C$ array.\qed
\end{proof}





\end{document}